\documentclass[11pt]{article}
\pdfoutput=1

\renewcommand{\Phi}{\phi}

\renewcommand{\Im}{{\rm Im \, }}

\usepackage[utf8]{inputenc}
\usepackage{geometry}
\usepackage{verbatim}
\usepackage{prettyref}
\usepackage{amsmath,mathtools}
\usepackage[normalem]{ulem}
\usepackage{graphicx}
\usepackage{setspace}
\usepackage{esint}
\usepackage{graphicx}
\usepackage{amsmath}
\usepackage{amsfonts}
\usepackage{amssymb}
\allowdisplaybreaks
\usepackage{physics}
\usepackage{microtype}
\usepackage{bbold}
\usepackage{cite}
\usepackage{tabu}
\usepackage{multirow}
\usepackage{hhline}
\usepackage{arydshln}
\usepackage{braket}

\usepackage{color}
\definecolor{darkgreen}{rgb}{0,0.5,0}
\definecolor{darkblue}{rgb}{0,0,0.6}
\definecolor{purple}{rgb}{0.4,.2,0.7}
\definecolor{orange}{rgb}{0.95, 0.5, 0.3}

\usepackage{cancel}

\usepackage[colorlinks=true,citecolor=darkgreen,linkcolor=darkblue,urlcolor=blue]{hyperref}

\numberwithin{equation}{section}

\newcommand\rref[1]{(\ref{#1})}

\def\be{\begin{equation}}
\def\ee{\end{equation}}
\def\bea{\begin{eqnarray}}
\def\eea{\end{eqnarray}}
\def\ba{\begin{align}}
\def\ea{\end{align}}

\makeatletter
\newcommand{\vast}{\bBigg@{4}}
\newcommand{\Vast}{\bBigg@{5}}
\makeatother

\usepackage{bbold}

\newcommand{\clo}{\mathcal{O}}
\newcommand{\clf}{\mathcal{V}}
\newcommand{\bbk}{\mathbb{K}}
\newcommand{\bbf}{\mathbb{F}}

\newcommand{\bbi}{\mathbb{1}}
\newcommand{\gb}{\Gamma_b}

\newcommand{\ord}{\mathcal{O}}
\newcommand{\clx}{X}
\newcommand{\cly}{Y}
\newcommand{\clw}{W}
\newcommand{\vd}{Q}

\newcommand{\den}{\rho_{\textnormal{vip}}}

\newcommand{\C}{C}

\newcommand{\eqspace}[1]{\phantom{.}\!\hspace{#1}}

\newcommand{\fker}[6]{
\mathbb{F}_{#1 #2}{\tiny\arraycolsep=0.2\arraycolsep\ensuremath{\begin{bmatrix}#4 & #3 \\ #5 & #6\end{bmatrix}}}
}

\begin{document}
\begin{spacing}{1.2}

~
\vskip5mm

~
\vskip5mm

\begin{flushright}
\hfill{\tt CERN-TH-2021-201}
\end{flushright}

\begin{center} {\Huge \textsc \bf OPE statistics from higher-point crossing}

\vskip10mm

Tarek Anous$^{1}$, Alexandre Belin$^{2}$, Jan de Boer$^{1}$ and Diego Liska$^{1}$
\vskip1em
{\it 1)  Institute for Theoretical Physics and $\Delta$-Institute for Theoretical Physics, University of Amsterdam, Science Park 904, 1098 XH Amsterdam, The Netherlands} \\
\vskip5mm
{\it 2) CERN, Theory Division,\\1 Esplanade des Particules, Gen\`{e}ve 23, CH-1211, Suisse \\}

\vskip5mm

\tt{ t.m.anous@uva.nl, a.belin@cern.ch, j.deboer@uva.nl, d.liska@uva.nl  }

\end{center}

\vskip10mm

\begin{abstract}

We present new asymptotic formulas for the distribution of OPE coefficients in conformal field theories. These formulas involve products of four or more coefficients and include light-light-heavy as well as heavy-heavy-heavy contributions. They are derived from crossing symmetry of the six and higher point functions on the plane and should be interpreted as non-Gaussianities in the statistical distribution of the OPE coefficients. We begin with a formula for arbitrary operator exchanges (not necessarily primary) valid in any dimension.  This is the first asymptotic formula constraining heavy-heavy-heavy OPE coefficients in $d>2$. For two-dimensional CFTs, we present refined asymptotic formulas stemming from exchanges of quasi-primaries as well as Virasoro primaries.

\

\end{abstract}

\pagebreak

\pagestyle{plain}

\setcounter{tocdepth}{3}
{}
\vfill
\tableofcontents

\section{Introduction}

Mapping out the space of all quantum field theories is one of the great ventures of modern physics. Conformal Field Theories (CFTs) are special points in this space of theories, since they correspond to fixed points of the renormalization group. One of the remarkable aspects of CFTs is that the dynamical information of the theory is packaged in a very concise way. At the local level, a CFT is specified by the spectrum of local operators (i.e. the list of scaling dimensions $\Delta_i$) and the OPE coefficients $C_{ijk}$ which control the operator algebra. This dynamical data is heavily constrained by consistency conditions: unitary, causality, crossing symmetry and, in two dimensions, modular invariance.\footnote{Modular invariance also constraints the theory in higher dimensions but it is harder to directly connect the constraints coming from modular invariance to the local data of the CFT. See \cite{Shaghoulian:2015kta,Belin:2016yll,Shaghoulian:2016gol,Belin:2018jtf}.} Exploiting these constraints to map out the space of CFTs is the venture of the bootstrap program \cite{Ferrara:1973yt,Polyakov:1974gs,Belavin:1984vu,Rattazzi:2008pe,Cardy:1991kr,Hellerman:2009bu}.

A powerful way to repackage some of the CFT constraints is in deriving asymptotic formulas for the CFT data. These formulas are constructed in the following way. Consider a correlation function or partition function and take a kinematical limit (it could be the OPE limit or a low-temperature limit): this will effectively project onto the lowest dimension operator compatible with the observable (usually the identity operator).  One then asks how this behaviour is reproduced in the cross-channel. In the cross-channel, it is never accomplished by a single operator, but rather comes from the collective contribution of many operators. The most famous example of this thinking is Cardy's formula for the density of states of two-dimensional CFTs \cite{cardyformula}
\be
\rho(\Delta)=e^{2\pi \sqrt{\frac{c}{3}\Delta}} \,.
\ee

While Cardy's formula relies on modular invariance, there are also many asymptotic formulas that one can derive from crossing symmetry. These formulas encode information about the averaged OPE coefficients, and the first formula of this kind appeared in \cite{Pappadopulo:2012jk} and reads\footnote{We will be more explicit about the nature of the average in the following section. Also note that this formula applies in any CFT (not only in $d=2$), such that the density of states is given by the appropriate density of states for the CFT at hand.}
\be \label{rattazziresult}
\overline{ |C_{L L {H}}|^2 }\sim \frac{ \Delta_H^{2\Delta_L-1}}{\rho(\Delta_H)} \,.
\ee
This formula comes from crossing invariance of the Euclidean four-point function on the plane and the sum is over all operators, including both primary and descendants. Since then, numerous formulas involving OPE coefficients have appeared \cite{Kraus:2016nwo,Das:2017vej,Cardy:2017qhl,Das:2017cnv,Qiao:2017xif,Mukhametzhanov:2018zja,Pal:2019zzr,Gobeil:2018fzy,Brehm:2018ipf,Romero-Bermudez:2018dim,Hikida:2018khg,Collier:2019weq,Belin:2021ryy}. The different formulas correspond to how many operators are heavy (H) and summed over, as opposed to those that are light (L) and fixed. They also depend on spacetime dimension, and whether we are summing over primaries or not. A summary of the known asymptotic formulas is given in Table \ref{tab:status}. Currently, there exists no asymptotic formula involving $C_{HHH}$ in $d>2$, and one of the goals of this paper is to fill this gap.\footnote{Very recently, an analysis of heavy-heavy-heavy OPE coefficients valid at large spin was carried out in \cite{Bercini:2020msp,Antunes:2021kmm}, which comes from studying correlation functions in the lightcone limit. In this paper, we will consider a different kinematic limit: the Euclidean OPE. This probes OPE coefficients of all spin.} 

\begin{table}[t]
\begin{center}
\begin{tabular}{lllll} 
\hline
                           & \multicolumn{2}{l}{$d=2$}                               & \multicolumn{2}{l}{$d>2$}                                \\ 
\hline
$C_{LLH}$                  & $\overline{|C_{LLH}|^2}$  & $\checkmark\phantom{\Big(}$ & $\overline{|C_{LLH}|^2}$  & $\checkmark\phantom{\Big(}$  \\ 
\hdashline[1pt/1pt]
\multirow{2}{*}{$C_{LHH}$} & $\overline{C_{LHH}}$      & $\checkmark\phantom{\Big(}$ & $\overline{C_{LHH}}$      & $\checkmark\phantom{\Big(}$  \\
                           & $\overline{|C_{LHH'}|^2}$ & $\checkmark\phantom{\Big(}$ & $\overline{|C_{LHH'}|^2}$ & $\checkmark\phantom{\Big(}$  \\ 
\hdashline[1pt/1pt]
$C_{HHH}$                  & $\overline{|C_{HHH}|^2}$  & $\checkmark\phantom{\Big(}$ & \multicolumn{2}{c}{?}                                    \\
\hline
\end{tabular}

\end{center}
\caption{Current status of the various known CFT asymptotic formulas for either light (L) or heavy (H) operators. It remains to constrain asymptotic formulas involving $C_{HHH}$ in $d>2$, which is one of the goals of the present paper.}
\label{tab:status}
\end{table}

Studying asymptotic formulas for OPE coefficients has become more relevant than ever due to the surge of interest in understanding the statistical distribution of the CFT data, in particular in the context of holography. On the gravitational side, this ties to the physics of wormholes which can capture operator statistics of the CFT \cite{Saad:2019pqd,Pollack:2020gfa, Belin:2020hea,Stanford:2020wkf,Blommaert:2020seb,Belin:2020jxr,Altland:2021rqn,Freivogel:2021ivu,Goto:2021mbt,Belin:2021ibv}. In \cite{Belin:2020hea}, a framework called the OPE Randomness Hypothesis (ORH) was proposed to systematically capture the statistical distribution of OPE coefficients. This hypothesis generalizes the ETH ansatz \cite{deutsch1991quantum,srednicki1994chaos} to CFTs and treats any OPE coefficient involving a heavy operator as a pseudo-random variable (a variable which is fixed in any given theory but that has a statistical distribution once sampled over enough OPE coefficients). 

From the point of view of the ORH, the asymptotic formulas constrain the moments of the distribution of OPE coefficients. The formulas given in Table \ref{tab:status} include one or two OPE coefficients, and thus represent the mean or variance of the distributions. Until recently, very little was known about the higher moments. In the case of the ETH ansatz, these moments are necessarily non-trivial, even though they are further suppressed in the entropy \cite{Foini:2018sdb} (see also \cite{PhysRevLett.123.230606,PhysRevLett.122.220601,Dymarsky:2018ccu,Richter:2020bkf,Wang:2021mtp}). It is natural that a similar statement should follow for the ORH. A first venture in this direction was conducted in \cite{Belin:2021ryy}, where higher moments of $C_{HHH}$ were obtained in two-dimensional CFTs exploiting modular invariance at genus $g>2$. In this paper, we will present other asymptotic formulas that complement the non-Gaussianity analysis for heavy OPE coefficients. These moments will contain both $C_{LLH}$ and $C_{HHH}$ OPE coefficients. The formulas will be obtained from the constraints of higher-point crossing on the plane.

\subsection{Summary of Results}

In this paper, we derive the following new asymptotic formulas for the densities $C_{LLH}^3C_{HHH}$ from crossing symmetry of the six-point function in the star channel
\bea \label{resultsintro}
\overline{C_{LLH}^3C_{HHH}}\Big|_{d\geq2, \ \textrm{all op}} &\sim&  \frac{\Delta_H^{6\Delta_L-3}}{\rho(\Delta_H)^3} \notag \\
\overline{C_{LLH}^3C_{HHH}}\Big|_{d=2, \ \textrm{quasi-prim}}&\sim& \frac{\Delta_{H}^{6\Delta_L-6}}{\rho(\Delta_H)^3} \\
\overline{C_{LLH}^3C_{HHH}}\Big|_{d=2, \ \textrm{Vir-prim}}&\sim& \left(\frac{3\sqrt{3}}{16}\right)^{3\Delta_H}\frac{\Delta_H^{6\Delta_L-\frac{19+11c}{36}}}{\den(\Delta_H)^{\frac{9}{4}}} \notag \,.
\eea
The notation $\sim$ will be made clearer in the following sections but for the first and third lines, it should be understood as meaning that the ratio of the true density and the result \rref{resultsintro} approaches a $\Delta_H$-independent number as $\Delta_H\to\infty$. For Virasoro primaries in $d=2$, we can even obtain an asymptotic formula valid for an arbitrary number of OPE coefficients\footnote{The asymptotic formula with $\eta=1$ (which comes from odd-point functions) is only applicable if $C_{LLL}\neq0$ (by $C_{LLL}$, we mean the OPE coefficient $C_{OOO}$ where $O$ is the external operator of the odd-point function). If $C_{LLL}=0$, then the formula does not apply and we were not able to derive an asymptotic formula in that case. A more in-depth discussion of this problem is given at the end of section \ref{sec4}. }
\begin{equation} \label{resultallpoint}
\overline{C_{LLH}^{m+2}C_{HHH}^mC_{HHL}^\eta}\Big|_{d=2, \ \textrm{Vir-prim}}\!\!\sim C_{LLL}^{\eta}\frac{3^{\; \frac{9m}{2}\Delta_H}}{16^{(1+2m)\Delta_H}}\;\frac{ \Delta_H^{\left(4+2m+\frac{\eta}{2}\right)\Delta_L-\frac{9(c+1)+2(5+c)m}{36}}}{\den(\Delta_H)^{ \frac{1}{4}\left(2+4\eta+7 m\right)}},
\end{equation} 
where $\den(\Delta_H)$ is the density of primary states at large conformal dimension
\begin{equation}
\den(\Delta) \sim e^{2\pi \sqrt{\frac{(c-1)}{3}\Delta}}.
\end{equation}
Equation \eqref{resultallpoint} was derived from the $(2m+4+\eta)$-point function, with $\eta$  being either zero or one. 

Our formulas are derived from crossing invariance of higher-point functions. The general strategy is always the same: We go to an OPE limit and keep the identity contribution, and use it to constrain the OPE density in the cross-channel. However, the technical analysis is slightly different in each case we consider. For the OPE density involving all operators (i.e. the first line of \rref{resultsintro}), we do not restrict the set of exchanged operators and can thus proceed directly by inspecting the crossing constraint. For quasi-primary exchanges in $d=2$, we are required to study conformal blocks of six-point functions. While these blocks are  known in closed form \cite{Fortin:2020yjz,Anous:2020vtw}, we obtain the asymptotic behavior directly from the differential equations obeyed by the conformal blocks, which is enough to extract the asymptotic formula we seek. Finally, for Virasoro primaries, we exploit a method first developed in \cite{Collier:2019weq} to extract OPE densities directly from the Virasoro crossing kernel which is known in closed form \cite{Ponsot:1999uf,Ponsot:2000mt}.

The asymptotic formulas we give are approximate in two aspects. First of all, the spectrum of a (compact) CFT is discrete and hence the OPE density, just like the density of states, is a sum of delta functions. To obtain smooth functions like \rref{resultsintro}, one must therefore smear the true microscopic density of states over a sufficiently large energy window. This is how our overline notation $\overline{\left(\cdot\right)}$ should be understood. More details are given in section \ref{sec2}. Second, the formulas are derived from crossing symmetry by going to an OPE limit and keeping the identity operator. Considering other exchanges will produce subleading corrections  in $\Delta_H$ to \rref{resultsintro}, which will depend on the gap to the lightest operator that contributes.\footnote{For the first two lines of \rref{resultsintro}, this will produce power-law corrections while it produces exponential corrections to the third line.}

A more precise formulation of both effects can be given using Tauberian theorems (see \cite{Qiao:2017xif,Mukhametzhanov:2018zja,Pal:2019zzr,Mukhametzhanov:2019pzy,Mukhametzhanov:2020swe,Das:2020uax}). It is worthwhile to note that the size of the window over which one should average is physically important. The asymptotic formulas derived here can apply to free field theories, but they will be realized very differently than in a chaotic CFT.\footnote{There is currently no clear-cut definition of chaos in CFTs, but for the purposes mentioned here, we could take a two-dimensional CFT with $c>1$ and no extended chiral algebra. For example, we expect the averaging windows to be parametrically smaller in such theories than in a product of free bosons.}  One way to diagnose this is to understand precisely how big of a window one should average over for the asymptotic formulas to apply. This is a theory-dependent statement, and goes beyond the scope of the present work. It is however important to emphasize that many combinations of OPE coefficients we consider are not positive-definite, contrary to previous asymptotic formulas (for example those for $|C_{LLH}|^2$ or $|C_{HHH}|^2$). Therefore, the fluctuations could be much larger and could impact the size of the averaging window. This is already the case for the torus one-point functions \cite{Kraus:2016nwo}.   

The paper is organized as follows, in section \ref{sec2}, we start by reviewing the Euclidean four-point crossing and derive a new asymptotic formula from six-point crossing which is valid in arbitrary dimensions. In section \ref{sec3}, we focus on $d=2$ CFTs and derive an asymptotic formula for quasi-primaries. In section \ref{sec4}, we use the Virasoro crossing kernel to derive an asymptotic formula for Virasoro primaries in $d=2$. We conclude with a summary and some open questions in section \ref{sec5}. Some details of the Virasoro crossing kernel are given in the appendix.

\section{Crossing Symmetry and Asymptotics for all Operators \label{sec2}}

In this section, we will start by reviewing the procedure of \cite{Pappadopulo:2012jk} to obtain the OPE density from the four-point crossing on the plane. We will use slightly different kinematics, which are well-tailored to be generalized to higher-point functions. We will then generalize the procedure to the six-point function. The results presented in this section will hold in arbitrary dimensions, but involve OPE densities for all operators, namely including both primary and descendants of arbitrary spin.

\subsection{Review of the four-point crossing}

\begin{figure}
    \centering
    \includegraphics[scale=0.6]{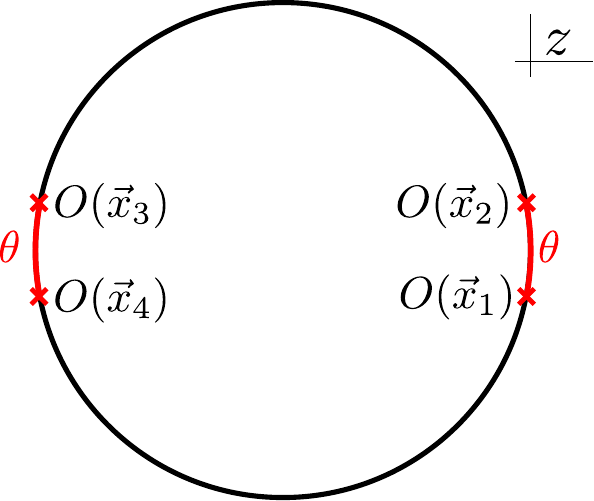}
    \caption{The kinematics considered for the four-point function in the $z$ plane. The black circle is the unit circle and the pairs of operators are inserted at an angle separation $\theta$, with one pair centered at 0 and one pair centered at $\pi$.}
    \label{4ptfig}
\end{figure}

In this section, we rederive the asymptotic formula for $|C_{LLH}|^2$ from the four-point function on the plane using slightly different kinematics. Consider the four-point function of identical scalars of dimension $\Delta_L$
\be
G_4=\braket{O(\vec{x}_1) O(\vec{x}_2) O(\vec{x}_3) O(\vec{x}_4)} \,.
\ee
The kinematics are as follows: Using conformal invariance, we can always map a four-point function to a two-dimensional plane. We will thus label points by their coordinate on the complex plane
\be
z_i=x^0_i + i x^1_i \quad \qquad x^a=0 \,, \quad 2\leq a \leq d-1 \,.
\ee
Instead of picking the four-points to be at $0,z,1,\infty$ as in \cite{Pappadopulo:2012jk}, we pick the four-points to be
\be
z_1 = e^{-i \frac{\theta}{2}}~, \qquad\qquad
z_2= e^{i \frac{\theta}{2}} ~, \qquad\qquad
z_3 = -e^{-i \frac{\theta}{2}} ~, \qquad\qquad
z_4= -e^{i \frac{\theta}{2}}  \,.
\ee
This is depicted in Figure \ref{4ptfig}. Naturally, because a four-point function only depends on one cross-ratio, there is no new physics here compared to \cite{Pappadopulo:2012jk}, but it will be more convenient to lift to higher-point functions where we will make restrictions on the cross-ratios.

We now study the OPE limit corresponding to taking $\theta\to0$, leading to:
\be \label{OPElimit4pt}
G_4=\frac{1}{|z_{12}|^{2\Delta_L}|z_{34}|^{2\Delta_L}}\left(1+\dots\right)\sim \frac{1}{\theta^{4\Delta_L}} \,.
\ee
The question set upon us is how to reproduce this singularity from the expansion in the cross-channel, namely the $1\leftrightarrow4$ and $2\leftrightarrow3$ OPE. 
{We will write this OPE explicitly as: 
\be
O(\vec{x}_1)O(\vec{x}_4)= \sum_{O_1} B_{OOO_1} |\vec{x}_{14}|^{\Delta_{O_1}-2\Delta_L} O_1(\vec{x}_4) \,,
\ee
and similarly for the remaining two operators. The operator $O_1$ appearing in the OPE can either be primary or descendant, but throughout we will drop spin structures. If $O_1$ is a primary, then $B_{OOO_1}$ is a standard three-point coefficient, but for descendants it will get dressed by various factors. In this OPE channel we have:
\begin{equation}
G_4\approx\sum_{\Delta_1,\Delta_2}\sum_{\substack{O_1\in\Delta_{O_1}=\Delta_1\\ O_2\in\Delta_{O_2}=\Delta_2}} \frac{B_{OOO_1}B_{OOO_2}}{{|z_{14}|^{2\Delta_L-\Delta_1}|z_{23}|^{2\Delta_L-\Delta_2}}}\langle O_1(\vec{x}_4)O_2(\vec{x}_2)\rangle~.
\end{equation}
The first sum in the above expression is over the dimensions of operators appearing in the OPE, and the second sum is over operators (primary or descendant) of a fixed conformal dimension. For fixed $\Delta_1$ and $\Delta_2$, we can diagonalize the matrix  $\langle O_1(\vec{x}_4)O_2(\vec{x}_2)\rangle$, and in this basis, which mixes the contribution of primaries and descendants, we have: 
\begin{equation}
G_4\approx\sum_{O_H} \frac{|C_{OOO_H}|^2}{{|z_{14}|^{2\Delta_L-\Delta_H}|z_{23}|^{2\Delta_L-\Delta_H}}}\langle O_H(\vec{x}_2)O_H(\vec{x}_4)\rangle~~,\label{eq:4ptallopsexpanded}
\end{equation}
where $C_{OOO_H}$ should be interpreted as the three-point coefficients in this new basis of operators. 
The approximate sign $\approx$ in the above expressions is coming from the fact that we dropped all spin dependence in these formulas. We expect that spin would not substantially change the results to come. Of course, to properly take advantage of conformal symmetry, we should really be summing over conformal blocks as we will do in the following two sections, this will both disentangle the effect of the descendants and spin. Here, however, we want to explicitly sum over all operators.\footnote{See \cite{Pappadopulo:2012jk} for more details on the relationship between $B_{OOO'}$ and $C_{OOO_H}$, in particular the comparison between equations (4.2) and (4.5) in their paper.}}

Our goal is now to reproduce the right pole in the small $\theta$ limit from the sum \eqref{eq:4ptallopsexpanded}, and to achieve this the OPE coefficients must therefore scale in the right way. We start by {assuming that $\langle O_H(\vec{x}_2)O_H(\vec{x}_4)\rangle$ coincides with the expression for primaries, and convert} the sum in \eqref{eq:4ptallopsexpanded} into an integral as follows: 
\begin{equation} \label{toinverseLaplace}
    G_4\approx\frac{1}{\left(2\cos\frac{\theta}{2}\right)^{4\Delta_L}}\int_0^\infty d\Delta_H\, \rho(\Delta_H) C^2(\Delta_H)\left(\cos\frac{\theta} {2}\right)^{2\Delta_H} \,,
\end{equation}
where the function $C^2(\Delta_H)$ is defined as follows: 
\begin{equation}
    C^2(\Delta_H)\equiv \sum_{O_{H'}}\frac{|C_{OOO_{H'}|^2}}{\rho(\Delta_{H'})}\delta(\Delta_H-\Delta_{H'})~,
\end{equation}
and $\rho(\Delta_H)$ is the density of states. Now for large $\Delta_H$, we will assume
\begin{equation}
    \lim_{\Delta_H\rightarrow\infty}C^2(\Delta_H)= \overline{ |C_{OOO_H}|^2}~,
\end{equation}
where the meaning of the overline, which will appear throughout our paper, is the following: 
\begin{equation}
 \int_0^{\Delta_H} d\Delta' \rho(\Delta') C^2(\Delta')\,f(\Delta') \approx\int_0^{\Delta_H} d\Delta' \rho(\Delta') \overline{|C_{OOO_{\Delta'}}|^2} \,f(\Delta')~, \qquad \Delta_H\rightarrow\infty~, \label{eq:overlinemeaning}
    \end{equation}
in the sense that the smooth function $\overline{|C_{OOO_\Delta}|^2}$ agrees with the exact set of three-point coefficients in a distributional sense, integrated against reasonable test functions $f(\Delta)$.\footnote{In the following sections where we perform a conformal block decomposition, the space of suitable test functions can be taken to be spanned by the conformal blocks.} 

While we could directly try to inverse Laplace transform \rref{toinverseLaplace}, we find it convenient to make an ansatz for the OPE density of the form
\be
\rho(\Delta_H) \overline{ |C_{OOO_H}|^2} \sim A_4^{-1}\Delta_H^{\gamma-1} \,,
\ee
we have to ensure:  
\be \label{sumtodo}
\frac{1}{\theta^{4\Delta_L}}\underset{\theta\rightarrow0}\sim\frac{A_4^{-1}}{\left(2\cos\frac{\theta}{2}\right)^{4\Delta_L}}\int d{\Delta_H} \Delta_H^{\gamma-1} \left(\cos\frac{\theta}{2}\right)^{2\Delta_H}\,.
\ee
This integral can be evaluated by saddle-point and the saddle is
\be
\Delta_H^*=\frac{1-\gamma}{2\log \left(\cos\frac{\theta}{2}\right)}\underset{\theta\to0}{\sim}\frac{4(\gamma-1)}{\theta^2}~.
\ee
This gives us an estimate of the scaling dimension which are contributing the most to the integral. We will come back to this later for the higher-point crossing. Performing the saddle-point integral and subsequently expanding for small $\theta$, we find
\be
G_4\sim \frac{\sqrt{\pi}2^{2\gamma-4\Delta_L+\frac{1}{2}}}{A_4}\times\frac{e^{1-\gamma}(\gamma-1)^{\gamma-\frac{1}{2}}}{\theta^{2\gamma}} \,.
\ee
Matching with \rref{OPElimit4pt}, we thus have
\be
\gamma=2\Delta_L \,,\quad\qquad A_4=e^{1-2\Delta_L}\sqrt{2\pi}\left(2\Delta_L-1\right)^{2\Delta_L-\frac{1}{2}}~,
\ee
and
\be
\rho(\Delta_H)\overline{ |C_{OOO_H}|^2} =\frac{ \Delta_H^{2\Delta_L-1}}{A_4(\Delta_L)} \,.
\ee
We can obtain a different estimate for the coefficient by noting that \rref{sumtodo} can be integrated without the need to resort to a saddle-point analysis. Performing this step, one finds
\be \label{finalresultrattazzi}
\overline{|C_{LLH}|^2}\equiv\overline{|C_{OOO_H}|^2} \sim \frac{\Delta_H^{2\Delta_L-1}}{\Gamma(2\Delta_L)\rho(\Delta_H)} \,,
\ee
which more closely resembles the result presented in \cite{Pappadopulo:2012jk}. Note that the saddle-point evaluation of the integral (with the one-loop determinant) gives a different coefficient $A_4$, but that it agrees with \rref{finalresultrattazzi} to a very good approximation for $\Delta_L\gtrsim 0.5$. Indeed, the scaling dimension of the light operator controls how peaked the integral is around its saddle.

We are now ready to study the six-point function.


\subsection{6-point crossing and the asymptotic formula for \texorpdfstring{$C_{LLH}^3 C_{HHH}$}{COOHcubedCHHH}}\label{sec:sixpointallops}

We will now work out a formula for the asymptotics of OPE coefficients $C_{LLH}^3 C_{HHH}$. Just like the formula of the previous section, it works in arbitrary dimension and includes contributions from all operators, whether primary or descendant, and of arbitrary spin. This formula will follow from crossing invariance of the six-point function on the plane. For two-dimensional CFTs, modular invariance at genus-two constrains the variance of the OPE coefficients $C_{HHH}$ but in higher dimensions, we believe that the six-point function is the simplest observable that probes $C_{HHH}$, and it does so through the combination $C_{LLH}^3 C_{HHH}$.\footnote{ The combination  $C_{LLH}^3 C_{HHH}$ has an interesting holographic interpretation. $C_{LLH}$ measures the probability of two particles colliding and forming a black hole \cite{Das:2017cnv}. The combination $C_{LLH}^3 C_{HHH}$ measures the probability of three pairs of particles forming a black hole, which in turn collide with one another. }

Moving to higher-point functions, one is immediately confronted with a complication. Conformal symmetry does not restrict higher-point functions to lie on a two-dimensional plane. At the level of asymptotic formulas, this is ultimately related to spin, and to the various ways to probe the spin dependence of OPE coefficients. In this paper, we will study the simplest possible asymptotic formula where we place all operators on a two-dimensional plane.

\begin{figure}
    \centering
    \includegraphics[scale=0.6]{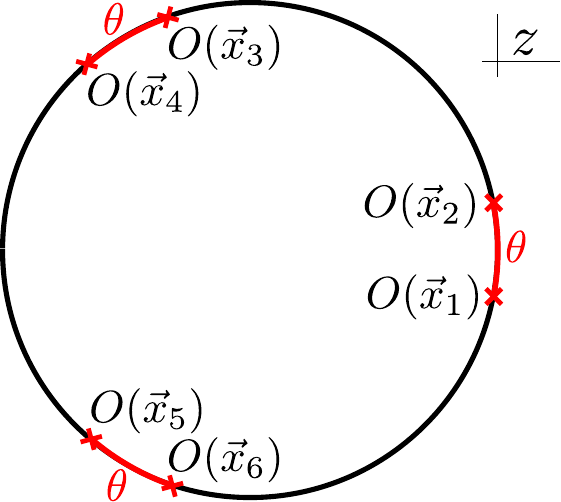}
    \caption{The kinematics considered for the six-point function in the $z$ plane. The black circle is the unit circle and the pairs of operators are inserted at an angle separation $\theta$, with one pair centered at 0, one pair centered at $2\pi/3$ and the third pair centered at $4\pi/3$.}
    \label{6ptfig}
\end{figure}

We now focus on the six-point function of identical scalar operators of dimension $\Delta_L$. We align all six operators on a two-dimensional plane (i.e. $x^a_i=0$ for all $a\geq2$ and for all $i$). We then place our operators on the unit circle, grouped up in pairs as follows 
\begin{align}
&z_1 = e^{-i \frac{\theta}{2}}~,& 
&z_3 = e^{-i \frac{\theta}{2}+\frac{2\pi i}{3}}~,& 
&z_5 = e^{-i \frac{\theta}{2}+\frac{4\pi i}{3}}~,&   \nonumber\\
&z_2= e^{i \frac{\theta}{2}}~,& 
&z_4= e^{i \frac{\theta}{2}+\frac{2\pi i}{3}}~,&
&z_6= e^{i \frac{\theta}{2}+\frac{4\pi i}{3}} \,.
\end{align}
This is depicted in Figure \ref{6ptfig}.
We again define
\begin{equation}
 G_6\equiv\braket{O(\vec{x}_1) O(\vec{x}_2) O(\vec{x}_3) O(\vec{x}_4) O(\vec{x}_5) O(\vec{x}_6)}   ~, 
\end{equation}
{}and once again, we consider the triple OPE limit, i.e. the limit $\theta\to0$, where 
\be \label{leadingdiv}
 G_6=\frac{1}{|z_{12}|^{2\Delta_L}|z_{34}|^{2\Delta_L}|z_{56}|^{2\Delta_L}}\left(1+\dots\right)\sim  \frac{1}{ \theta^{6\Delta_L}} \,.
\ee
We will now ask how this singularity is reproduced in another convergent  OPE channel, namely the fusions $1\leftrightarrow6, 2\leftrightarrow3, 4\leftrightarrow5$. Again writing the OPE explicitly as: 
\be
O(\vec{x}_1)O(\vec{x}_6)= \sum_{O_1} B_{OOO_1} |\vec{x}_{16}|^{\Delta_H-2\Delta_L} O_1(\vec{x}_1) \,,
\ee
and similarly for the other pairs, we are left with three operators sitting at $x_1, x_3$ and $x_5$. Their three-point function reads\footnote{For now, we assume that the three operators are scalar primaries.}
\be
\braket{O_{1}(\vec{x}_1)O_{2}(\vec{x}_3)O_{3}(\vec{x}_5)}=\frac{B_{O_{1}O_{2}O_{3}}}{3^{\frac{\Delta_{1}+\Delta_{2}+\Delta_{3}}{2}}} \,.
\ee
Putting everything together, we find:
\begin{align}
G_6&=\sum_{O_{1},O_{2},O_{3}}\frac{B_{OOO_1}B_{OOO_2}B_{OOO_3}}{ |x_{16}|^{2\Delta_L-\Delta_1}|x_{23}|^{2\Delta_L-\Delta_2}|x_{45}|^{2\Delta_L-\Delta_3}}\braket{O_{1}(\vec{x}_1)O_{2}(\vec{x}_3)O_{3}(\vec{x}_5)}\nonumber\\
    &=\frac{1}{\left[2\cos\left(\frac{\theta}{2}+\frac{\pi}{6}\right)\right]^{6\Delta_L}}\sum_{O_{1},O_{2},O_{3}}C_{LLH}^3C_{HHH}\left[\frac{2}{\sqrt{3}}\cos\left(\frac{\theta}{2}+\frac{\pi}{6}\right)\right]^{\Delta_{1}+\Delta_{2}+\Delta_{3}}\label{crosschannel6pt}
\end{align}
where we have defined
\be
C_{LLH}^3C_{HHH} \equiv B_{OOO_{1}}B_{OOO_{2}} B_{OOO_{3}}  B_{O_{1}O_{2}O_{3}} \,.
\ee
The final formula we will derive will be for $C_{LLH}^3 C_{HHH}$ and it should be interpreted as a density for  OPE coefficients of primary operators, suitably convoluted with three sums over descendants.

Proceeding again by converting the sum into an integral: 
\begin{multline}\label{integral6pt}
   G_6=\frac{1}{\left[2\cos\left(\frac{\theta}{2}+\frac{\pi}{6}\right)\right]^{6\Delta_L}}\int d\Delta_1\, d\Delta_2\, d\Delta_3\,\rho(\Delta_1)\rho(\Delta_2)\rho(\Delta_3) \\\times \vd_{OOO}(\Delta_1,\Delta_2,\Delta_3)\left[\frac{2}{\sqrt{3}}\cos\left(\frac{\theta}{2}+\frac{\pi}{6}\right)\right]^{\Delta_{1}+\Delta_{2}+\Delta_{3}}~,
\end{multline}
where
\begin{align}
    \vd_{OOO}(\Delta_1,\Delta_2,\Delta_3)\equiv&\nonumber\\&\sum_{O_{1'},O_{2'},O_{3'}}\frac{\,B_{OOO_{1'}}B_{OOO_{2'}}B_{OOO_{3'}}B_{O_{1'}O_{2'}O_{3'}}}{\prod_{i'=1}^3\rho(\Delta_{i'})}\prod_{i=i'=1}^3\delta(\Delta_i-\Delta_{i'})~, 
\end{align}
and the OPE density must be determined in order to reproduce the correct pole \eqref{leadingdiv}. Assuming that
\begin{equation}
    \lim_{\Delta_1,\Delta_2,\Delta_3\rightarrow\infty} \vd_{OOO}(\Delta_1,\Delta_2,\Delta_3)=\overline{C_{LLH}^3C_{HHH}}~,
\end{equation}
we now consider the following ansatz for the OPE coefficients (which again could be derived from the appropriate inverse Laplace transforms):
\be
\rho(\Delta_{1})\rho(\Delta_{2})\rho(\Delta_{3}) \overline{C_{LLH}^3 C_{HHH}} \sim A_6^{-1}(\Delta_1)^{\gamma_1-1} (\Delta_2)^{\gamma_2-1}(\Delta_3)^{\gamma_3-1} \,,
\ee
and we will determine $\gamma_i$ by consistency with \rref{leadingdiv}. 
We can now evaluate the three integrals by saddle-point, which yields the saddle-point value of the conformal dimension at:
\be
\Delta_i^* = \frac{1-\gamma_i}{ \log\left[\frac{2}{\sqrt{3}}\cos\left(\frac{\theta}{2}+\frac{\pi}{6}\right)\right]}\underset{\theta\to0}{\sim} \frac{2\sqrt{3}(\gamma_i-1)}{\theta} ~. 
\ee
Evaluating the full integral \rref{integral6pt} directly, we obtain, in the $\theta\to0$ limit:
\begin{equation}
    G_6=\frac{1}{\theta^{6\Delta_L}}=\frac{A_6^{-1}}{\theta^{\gamma_1+\gamma_2+\gamma_3}}2^{\gamma_1+\gamma_2+\gamma_3}3^{\frac{\gamma_1+\gamma_2+\gamma_3}{2}-3\Delta_L}\Gamma(\gamma_1)\Gamma(\gamma_2)\Gamma(\gamma_3) \,,
\end{equation}
which suggests: 
\begin{equation}
    \gamma_1=\gamma_2=\gamma_3=2\Delta_L~,\qquad A_6=\left[2^{2\Delta_L}\Gamma(2\Delta_L)\right]^3~.
\end{equation}
If we now take $\Delta_1=\Delta_2=\Delta_3=\Delta_H$, we have:
\be \label{finalresultanyd}
\overline{ C_{LLH}^3 C_{HHH}}  \sim \frac{\Delta_H^{6\Delta_L-3}}{2^{6\Delta_L}\Gamma(2\Delta_L)^3\rho(\Delta_H)^3} \,.
\ee
We believe this to be the first asymptotic formula for an OPE coefficient with three heavy operators in a CFT in $d>2$.\footnote{A different formula valid at large spin has recently appeared in \cite{Bercini:2020msp,Antunes:2021kmm}.} 

Note that it should be interpreted as a non-Gaussianity in the four-point moment of three $C_{LLH}$ and one $C_{HHH}$ OPE coefficient. To see that there is some non-Gaussianity at play, consider the following expression
\be \label{ratioalld}
\frac{\overline{ C_{LLH}^3 C_{HHH}}}{\left[\overline{|C_{LLH}|^2}\right]^{3/2}} \sim \frac{ \Delta_H^{3\Delta_L-3/2} }{\rho(\Delta_H)^{3/2}} \,.
\ee
If the moments are factorized, one could be tempted to attribute the result \rref{ratioalld} solely to $C_{HHH}$. But this is clearly impossible: The OPE coefficients $C_{HHH}$ are defined irrespective of the light operators so their density cannot depend on $\Delta_L$. Therefore, it is clear that the moments do not factorize and we have some non-Gaussianity at play. 

This concludes our analysis in arbitrary dimensions. In what follows, we will restrict to two-dimensional conformal field theories. In that case, we will be able to get more refined information by separating the contribution of primaries and descendants. We will be able to obtain expressions both for quasi-primary operators, and for Virasoro primary operators.


\section{Quasi-Primaries and Six-point Crossing \label{sec3}}

We will now restrict our attention to two-dimensional conformal field theories. In this section, we study the quasi-primary decomposition of the six-point function in the star-channel and derive an expression for the conformal blocks in the asymptotically heavy limit. We use these expressions to obtain the OPE density for quasi-primary operators.\footnote{We will see in the following section that we can derive OPE densities directly for Virasoro primaries, which is in general a more useful and refined description of the dynamics. However, it is also of interest to study the effect of descendants on the asymptotic formulas, which can be obtained by comparing the formulas we derive here with those of the following section. Moreover, we believe that the calculations of this section will be a useful stepping stone for extracting primary densities from the six-point function in $d>2$, which we leave for future work.}

\subsection{Review of Qiao+Rychkov's argument for the four-point function}\label{sec:d24pt}
We start by reviewing an argument given in \cite{Qiao:2017xif}, suitably modified to facilitate a generalization to higher points. 
We start with the correlation function of four identical operators of dimension $\Delta_L=h_L+\bar{h}_L$ and spin $s=h_L-\bar{h}_L$ in a 2d CFT:
\begin{equation}
	G_4(z_1,\bar{z}_1,z_2,\bar{z}_2,z_3,\bar{z}_3,z_4,\bar{z}_4)\equiv\left\langle O(z_1,\bar{z}_1)O(z_2,\bar{z}_2)O(z_3,\bar{z}_3)O(z_4,\bar{z}_4)\right\rangle~.
\end{equation}
The $z_i$ are holomorphic coordinates on the complex plane and $\bar{z}_i$ are their complex conjugates. As usual, we will treat $z_i$ and $\bar{z}_i$ as independent coordinates. By conformal invariance, $G$ is fixed to take the following form: 
\begin{equation}
	G_4=\frac{1}{z_{12}^{2h_L}z_{34}^{2h_L}}\frac{1}{\bar{z}_{12}^{2\bar{h}_L}\bar{z}_{34}^{2\bar{h}_L}}F_4(z,\bar{z})~,\label{eq:eqGdef}
\end{equation}
with 
\begin{equation}
	z\equiv\frac{(z_1-z_2)(z_3-z_4)}{(z_1-z_3)(z_2-z_4)}~, \qquad \bar{z}\equiv\frac{(\bar{z}_1-\bar{z}_2)(\bar{z}_3-\bar{z}_4)}{(\bar{z}_1-\bar{z}_3)(\bar{z}_2-\bar{z}_4)}~. 
\end{equation}
Moreover, performing the operator product expansion in the limit $z_1\rightarrow z_2$ and $z_3\rightarrow z_4$, we can further decompose $F$ as follows: 
\begin{align}
	F_4(z,\bar z)&=\sum_{O_H}\begin{gathered}\includegraphics[height=1.8cm]{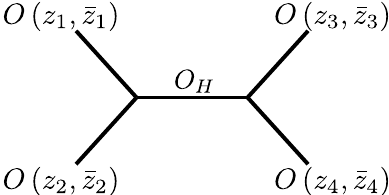}\end{gathered}\nonumber\\
	&=\int_0^\infty dh_H\, \int_0^\infty d\bar{h}_H\,\rho(h_H,\bar{h}_H)\,C^2(h_H,\bar{h}_H) \, \mathcal{F}_{h_H}(z){\mathcal{F}}_{\bar{h}_H}(\bar{z})~,\label{eq:OPE4pt}
\end{align}
where we have defined
\begin{equation}\label{eq:defcsquared}
	\,C^2(h_H,\bar{h}_H)\equiv \sum_{O_{H'}}\frac{C^2_{OOO_{H'}}}{\rho(h_{H'},\bar{h}_{H'})}\delta(h_H-h_{H'})\delta(\bar{h}_H-\bar{h}_{H'})~,
\end{equation}
and the functions $\mathcal{F}_{h_H}(z)$ are known as \emph{global conformal blocks} and satisfy a Casimir equation of the global conformal symmetry:
\begin{equation}
	z^2\left[(1-z)\partial_z^2-\partial_z\right]\mathcal{F}_{h}(z)=h(h-1)\mathcal{F}_{h}(z)~,\label{eq:cas4pt}
\end{equation}
with known solution 
\begin{equation}
	\mathcal{F}_{h}=z^{h} {}_2F_1(h,h,2h,z)~.\label{eq:hypergeometricconfblock}
\end{equation}
The $C_{OOO_H}$ are the three-point coefficients labeling the allowed fusions of quasi-primaries $O\times O\rightarrow O_H$ in the theory of interest and $\rho(h,\bar{h})$ is the density of states. For now, we are summing over all quasi-primary exchanged operators $O_H$, but the meaning of the subscript will become clear in what follows.

As written, it is clear that the correlation function \eqref{eq:eqGdef} has the appropriate singularity in the OPE limit $z_2\rightarrow z_1$ $(z\rightarrow 0)$. But the expansion \eqref{eq:OPE4pt} should also capture the cross-channel OPE singularity in the limit $z_2\rightarrow z_3$ $(z\rightarrow 1)$~. In order to reproduce the cross channel OPE singularity, it is sufficient that the function $F$ behave as:
\begin{equation}\label{eq:Flimits}
	F(z,\bar{z})\underset{z\rightarrow 1}{\sim} (1-z)^{-2h_L} \bar{H}(\bar{z} ) ~,\qquad\qquad F(z,\bar{z})\underset{\bar{z}\rightarrow 1}{\sim} (1-\bar{z})^{-2\bar{h}_L} {H}(z )~, 
\end{equation}
as one can easily check.\footnote{In more detail, this behavior would ensure that as $z_2\rightarrow z_3$, 
\begin{equation}
\nonumber
	F\sim \left[\frac{(z_1-z_4)(z_2-z_3)}{(z_1-z_3)(z_3-z_4)}\right]^{-2h_L}\bar{H}(\bar{z})~. 
\end{equation}
We leave it up to the reader to check that this combines with the prefactor in \eqref{eq:eqGdef} to give the correct cross-channel singularity. } 

However, since the conformal blocks are fixed, known (hypergeometric) functions with a $\log 1-z$ singularity in the limit $z\rightarrow 1$, this desired behavior must be the result of  a cumulative effect stemming from summing over the various blocks. This, in turn acts as a constraint on the asymptotic form of the density of three-point coefficients $C^2(h_H,\bar{h}_H)$ for large $h_H$ and $\bar{h}_H$. We proceed to show how this works. In what follows, we will focus on the holomorphic dependence of the correlation function. The argument for the anti-holomorphic dependence will follow essentially unchanged. 

We start with the defining equation for the blocks \eqref{eq:cas4pt} and change coordinates $z\rightarrow 1-y/h_H^2$:
\begin{equation}
	0=\frac{\left(h_H^2-y\right)^2}{h_H^2}\left[y\partial_y^2+\partial_y\right]\mathcal{F}_{h_H}(y)-h_H(h_H-1)\mathcal{F}_{h_H}(y)~.
\end{equation}
To zoom in on the cross-channel limit $(z\rightarrow 1)$, we take $h_H\rightarrow\infty$ (the exchanged operator is `heavy') and $y\sim O(1)$ and work to leading order:
\begin{equation}
	0=h_H^2\left[y\partial_y^2+\partial_y-1\right]\mathcal{F}_{h_H}(y)+O(h_H)~ \,. 
\end{equation}
We recognize this leading-order ODE as Bessel's differential equation, with solution:
\begin{equation}
	\mathcal{F}_{h_H}(y)\approx c(h_H)K_0(2\sqrt{y})~,
\end{equation}
where we have chosen the $K_0$ (rather than an admixture including $I_0$) solution in order to mimic the hypergeometric-function's behavior as a monotonically increasing function as we approach $z=1$ ($y=0$) from below (above), with a logarithmic singularity at $z=1$. 
Thus we have shown that for $z=1-x$, with $h_H\rightarrow\infty$ and $h_H^2 x\sim O(1)$ :
\begin{equation}
	\mathcal{F}_{h_H}(1-x)\approx c(h_H)K_0(2h_H\sqrt{x})~,  
\end{equation}
up to an undetermined coefficient $c(h_H)$.\footnote{Since the conformal block \eqref{eq:hypergeometricconfblock} has a fixed leading coefficient $\mathcal{F}_h\approx z^h\left(1+\frac{hz}{2}+\dots\right)$ in the $z\sim 0$ expansion, \cite{Qiao:2017xif} are able to fix \begin{equation} \nonumber
	c(h_H)=\frac{\Gamma(2h_H)}{\Gamma(h_H)^2}\approx 4^{h_H}\sqrt{\frac{h_H}{\pi}}~.
\end{equation}
Using only the asymptotics of the differential equation, we are unable to similarly fix the coefficient $c(h_H)$. However, this coefficient is tied to an overall normalization of the basis vectors, and different choices lead to different $c(h_H)$. } In order to reproduce the desired behavior, we need to assume two things, first:
\begin{equation}
	\lim_{h_H,\bar{h}_H\rightarrow\infty}C^2(h_H,\bar{h}_H)\equiv\overline{C_{OO O_H}^2}=\overline{\mathcal{C}_{OO O_H}^2}\times\overline{\mathcal{C}_{\bar{O}\bar{O} \bar{O}_H}^2} \,,
\end{equation}
meaning the asymptotic dependence of the averaged-squared-three-point coefficients factorizes. Secondly, we need: 
\begin{equation}
	\int_0^\infty dh_H \,\rho(h_H)\overline{\mathcal{C}_{OOO_H}^2}c(h_H)K_0(2h_H\sqrt{1-z})\underset{z\rightarrow1}{\sim}(1-z)^{-2h_L}~. 
\end{equation}
In these formulas, we have written overlines such as $\overline{C_{OOO_H}^2}$ in order to indicate that this is a constraint on the \emph{average} density of three-point coefficients at large-$h_H$, in the sense described in \eqref{eq:overlinemeaning}. Moreover, we have used the fact that the Cardy density factorizes asymptotically between holomorphic and anti-holomorphic sectors.  In order to proceed, we will assume the asymptotic form 
\begin{equation}
	\rho(h_H)\overline{\mathcal{C}_{OOO_H}^2}c(h_H)\underset{h_H\rightarrow\infty}\sim A_4^{-1} h_H^{\gamma-1} \,,
\end{equation}
for some yet undetermined $A_4$ and $\gamma$~. These, in turn, will be determined by demanding
\begin{equation}
	(1-z)^{-2h_L}=A_4^{-1}\int_0^\infty dh_H\,   h_H^{\gamma-1}K_0(2 h_H\sqrt{1-z})=(1-z)^{-\gamma/2}\frac{\Gamma\left(\frac{\gamma}{2}\right)^2}{4A_4}~ \,,
\end{equation}
which gives
\begin{equation}
	\gamma = 4h_L~,\qquad A_4=\frac{\Gamma\left({2h_L}\right)^2}{4}~. \label{eq:paramsintoverK}
\end{equation}
A more naive way to perform the above calculation, which readily generalizes to the higher point case, is to only use the asymptotic form of the Bessel function \cite{abramowitz1988handbook}:
\begin{equation}\label{eq:Kasymptotics}
	K_\nu(u)\underset{u\rightarrow\infty}{\approx}\sqrt{\frac{\pi}{2u}}e^{-u}\left[1-\frac{4\nu^2-1}{8u}+\frac{(4\nu^2-1)(4\nu^2-9)}{128u^2}+\dots\right]
\end{equation} 
which means we need: 
\begin{equation}
	(1-z)^{-2h_L}\approx A_4^{-1}(1-z)^{-1/4}\int_0^\infty dh_H\,   h_H^{\gamma-1}e^{-2h_H\sqrt{1-z}}\sqrt{\frac{\pi}{4h_H}}~. 
\end{equation}
We can estimate the above integral via a saddle point approximation (sensitive to the large $h_H$ part of the integral),  without forgetting the fluctuation integral around the saddle. This gives
\begin{equation}
	\gamma=4h_L~, \qquad A_4\approx2^{4h_L-2}\pi e^{-4h_L+\frac{3}{2}}\left(h_L-\frac{3}{8}\right)^{4h_L-1}\label{eq:4ptgammaA}~.
\end{equation}
This approximation for $A_4$ is not quite the same as Stirling's approximation for \eqref{eq:paramsintoverK}. However, one can check by plotting that, for $h_L\gtrsim 0.676$, this approximation upper bounds $\Gamma(2h_L)^2/4$~.
The important lesson is that
\begin{equation}
	\rho(h_H)\overline{\mathcal{C}_{OOO_H}^2}c(h_H)\underset{h_H\rightarrow\infty}\sim A_4(h_L)^{-1} h_H^{4h_L-1}~,
\end{equation}
thus reproducing the result in \cite{Qiao:2017xif}. The last step is to multiply this result by the anti-holomorphic dependence, meaning we have
\begin{equation}
    \boxed{\rho(h_H)\rho(\bar{h}_H)\overline{C_{OOO_H}^2}c(h_H)c(\bar{h}_H)\underset{h_H,\bar{h_H}\rightarrow\infty}\sim  \frac{h_H^{4h_L-1}\bar{h}_H^{4\bar{h}_L-1}}{A_4(h_L)A_4(\bar{h}_L)}}~.
\end{equation}

\subsection{Six-point analysis}
Now we consider 2d CFT correlation functions of six operators, grouped in pairs: 
\begin{multline}\label{eq:gencorr}
	G_6(x_1,\bar{x}_1,x_2,\bar{x}_2,y_1,\bar{y}_1,y_2,\bar{y}_2,w_1,\bar{w}_1,w_2,\bar{w}_2)=\\\langle X(x_1,\bar{x}_1)X(x_2,\bar{x}_2) Y(y_1,\bar{y}_1)Y(y_2,\bar{y}_2)W(w_1,\bar{w}_1)W(w_2,\bar{w}_2)\rangle~.
\end{multline}
The operators have dimensions $\Delta_X=h_X+\bar{h}_X$, $\Delta_Y=h_Y+\bar{h}_Y$ and  $\Delta_W=h_W+\bar{h}_W$ and spin $s_X=h_X-\bar{h}_X$, $s_Y=h_Y-\bar{h}_Y$ and  $s_W=h_W-\bar{h}_W$~. 

We can express the above correlation function as follows 
\begin{equation}
 	G_6=\frac{1}{x_{12}^{2h_X}y_{12}^{2h_Y}w_{12}^{2h_W}}\frac{1}{\bar{x}_{12}^{2\bar{h}_X}\bar{y}_{12}^{2\bar{h}_Y}\bar{w}_{12}^{2\bar{h}_W}}F_6(z,u,v,\bar{z},\bar{u},\bar{v})\label{eq:eqG6def}
 \end{equation} 
 where the function $F_6$ only depends on the following conformally invariant cross-ratios
 \begin{equation}
\label{eq:crossDef}
   z = \frac{(x_1-y_1)(y_2-x_2)}{(x_1-y_2)(y_1-x_2)} \,, \qquad u = \frac{(x_1-y_1)(y_2-w_1)}{(x_1-y_2)(y_1-w_1)} \,, \qquad v =  \frac{(x_1-y_1)(y_2-w_2)}{(x_1-y_2)(y_1-w_2)}\,,
\end{equation}
and their anti-holomorphic counterparts. As was the case before, associativity of the operator algebra allows for a decomposition of the function $F$ into an infinite sum over intermediate exchanges of operators. The functions labeling individual exchanges are known as \emph{conformal blocks} and by holomorphy of 2d CFT, they decompose into a product of a holomorphic function and an anti-holomorphic function. At six points, there are several choices for how these functions can be broken up, but we will choose the \emph{star} decomposition. In summary, we may write: 
\begin{align}\label{eq:decomp}
	F_6(z,u,v,\bar{z},\bar{u},\bar{v})&=\sum_{O_1,O_2,O_3}\begin{gathered}\includegraphics[height=3cm]{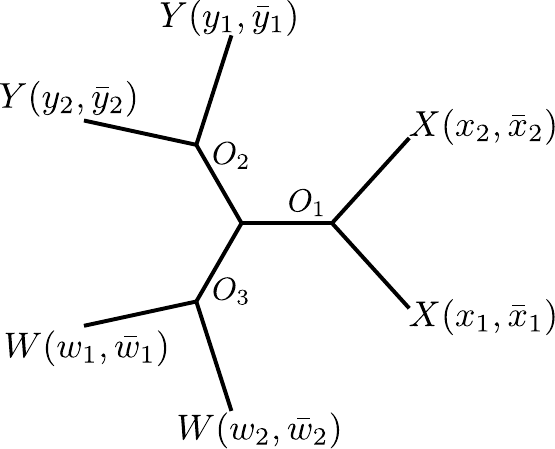}\end{gathered}\nonumber\\
	&=\int dh_1 dh_2 dh_3\,\int d\bar{h}_1 d\bar{h}_2 d\bar{h}_3\,\rho(h_1,\bar{h}_1)\rho(h_2,\bar{h}_2)\rho(h_3,\bar{h}_3)\,\nonumber\\
	&\quad\quad\quad\quad\times \vd_{WXY}(h_1,h_2,h_3,\bar{h}_1,\bar{h}_2,\bar{h}_3)\mathcal{F}_{h_1h_2h_3}(z,u,v){\mathcal{F}}_{\bar{h}_1\bar{h}_2\bar{h}_3}(\bar{z},\bar{u},\bar{v})~, 
\end{align}
where
\begin{multline}
	\vd_{WXY}(h_1,h_2,h_3,\bar{h}_1,\bar{h}_2,\bar{h}_3)\equiv\\\sum_{O_{1'},O_{2'},O_{3'}}\frac{\,C_{XXO_{1'}}C_{YYO_{2'}}C_{WWO_{3'}}C_{O_{1'}O_{2'}O_{3'}}}{\prod_{i'=1}^3\rho(h_{i'},\bar{h}_{i'})}\prod_{i=i'=1}^3\delta(h_i-h_{i'})\delta(\bar{h}_i-\bar{h}_{i'})~, 
\end{multline}
and once again the $C_{ijk}$ are the three-point coefficients of the theory. 
Expression \eqref{eq:eqG6def} makes clear that this expansion captures the OPE singularity as $x_1\rightarrow x_2$ ($z\rightarrow1$),  $w_1\rightarrow w_2$ ($u\rightarrow v$) and $y_1\rightarrow y_2$ ($u,v,z\rightarrow 1$).

Let us now restrict to the case where $X,Y,W=O$ are the same operator, meaning we take
\begin{equation}
 	h_X=h_Y=h_W\equiv h_L~, \qquad \bar{h}_X=\bar{h}_Y=\bar{h}_W\equiv \bar{h}_{L}~.
 \end{equation} 
 What is required of the density of OPE coefficients such that we reproduce the cross channel singularities as $x_2\rightarrow y_1$ ($z\rightarrow\infty$), $y_2\rightarrow w_1$ ($u\rightarrow 0$), and $w_2\rightarrow x_1$ ($v\rightarrow 1$)? We will use the fact \cite{Anous:2020vtw} that the six-point blocks in this channel obey the following Casimir equations: 
\begin{align}
&\left[-(b-c)^2\partial_b\partial_c-h_1(h_1-1)\right]\mathcal{F}_{h_1h_2h_3}(z,u,v)=0\label{eq:cas1}\\
&\left[(1-z)^2\left\lbrace \partial_z z\partial_z+\left(u\,\partial_u+v\,\partial_v\right)\partial_z\right\rbrace-h_2(h_2-1)\right]\mathcal{F}_{h_1h_2h_3}(z,u,v)=0\label{eq:cas2}\\
&\left[-(u-v)^2\partial_u\partial_v-h_3(h_3-1)\right]\mathcal{F}_{h_1h_2h_3}(z,u,v)=0\label{eq:cas3} \,,
\end{align}
with  $b\equiv\frac{u}{v}\frac{1-v}{1-u}$ and $c\equiv\frac{1-v}{1-u}$. For arbitrary $(h_1,h_2,h_3)$ these PDEs are difficult to solve, but as in the four-point case, we only need their asymptotic form.
We will approach the cross-channel limit in the following way: We first change to the following coordinates: 
\begin{equation}
 	z\rightarrow h_2 z' ~, \qquad u\rightarrow u'/h_1~, \qquad v\rightarrow 1-v'/h_3
 \end{equation} 
 and will simultaneously scale $(h_1,h_2,h_3)\rightarrow\infty$ while keeping the primed coordinates fixed and $O(1)$. This turns out to be the simplest scaling that produces non-trivial differential equations.  The leading equations become: 
 \begin{align}
 	0&=h_1\left[h_2z'^2\partial_{z'}\partial_{u'}+h_1\right]\mathcal{F}_{h_1h_2h_3} +O(h^1)\\
 	0&=h_2\left[h_3z'^2\partial_{z'}\partial_{v'}+h_2\right]\mathcal{F}_{h_1h_2h_3} +O(h^1)\\
 	0&=h_3\,\left[~-h_1\partial_{u'}\partial_{v'}+h_3\right]\mathcal{F}_{h_1h_2h_3} +O(h^1) \,,
 \end{align}
 and it is easy to check that all three of these equations are satisfied by
 \begin{equation}
 	\mathcal{F}_{h_1h_2h_3}\approx c(h_1,h_2,h_3)e^{-\frac{h_1}{h_3z'}-\frac{h_3u'}{h_2}-\frac{h_2v'}{h_1}}=c(h_1,h_2,h_3)e^{-\frac{h_1h_2}{h_3z}-\frac{h_1 h_3u}{h_2}-\frac{h_2h_3}{h_1}(1-v)}~, 
 \end{equation}
 up to an undetermined coefficient of $c(h_1,h_2,h_3)$~. Recall that the cross channel limit is $z\rightarrow\infty$, $u\rightarrow 0$, and $v\rightarrow 1$, or, if we define: 
 \begin{equation}
 	z\equiv1/x~,\qquad\qquad v\equiv 1-y~, 
 \end{equation}
 then the cross-channel limit is $(x,u,y)\rightarrow (0,0,0)$. In order to reproduce the cross-channel singularity, in complete analogy to \eqref{eq:Flimits}, we need
 \begin{equation}
 	F(x,u,y,\bar{z},\bar{u},\bar{v})\underset{(x,u,y)\rightarrow0}{\sim}x^{-2h_L}u^{-2h_L}y^{-2h_L} H(\bar{z},\bar{u},\bar{v})
 \end{equation}
 and similarly for the anti-holomorphic dependence in order to reproduce the cross-channel singularity in the anti-holomorphic variables. 
As in the four-point case we will assume a factorization for large conformal dimension: 
\begin{align}\nonumber 
	&\lim_{h_1,h_2,h_3,\bar{h}_1,\bar{h}_2,\bar{h}_3\rightarrow\infty}\vd_{OOO}(h_1,h_2,h_3,\bar{h}_1,\bar{h}_2,\bar{h}_3)\\\nonumber & \phantom{****.} \equiv \overline{C_{OO O_1}C_{OO O_2}C_{OO O_3}C_{O_1O_2O_3}} \\&\phantom{****.} =\overline{\mathcal{C}_{OO O_1}\mathcal{C}_{OO O_2}\mathcal{C}_{OO O_3}\mathcal{C}_{O_1O_2O_3}} \times \overline{\mathcal{C}_{\bar O\bar O \bar{O}_1}\mathcal{C}_{\bar O\bar O \bar{O}_2}\mathcal{C}_{\bar O\bar O \bar{O}_3}C_{\bar{O}_1\bar{O}_2\bar{O}_3}}~,\label{eq:globallargeHlimit}
\end{align}
meaning that we again assume factorization of the asymptotics between the holomorphic and anti-holomorphic sectors. We will proceed by assuming factorization, but since we will obtain the correct answer, this is enough to show that the factorization must occur. Indeed, we could prove it  instead by inverse Laplace transforming the OPE pole.

Note that this is a statement about the average of \emph{this particular product of OPE coefficients}, implying that it gives us access to the non-Gaussianities of the distribution over OPE coefficients. We will now reinterpret the cross-channel singularity as the following statement: 
 \begin{multline}
 	x^{-2h_L}u^{-2h_L}y^{-2h_L}=\int dh_1 dh_2 dh_3\,\rho(h_1)\rho(h_2)\rho(h_3)\\ \times\overline{\mathcal{C}_{OO O_1}\mathcal{C}_{OO O_2}\mathcal{C}_{OO O_3}\mathcal{C}_{O_1O_2O_3}}c(h_1,h_2,h_3)e^{-\frac{h_1h_2}{h_3}x-\frac{h_1 h_3}{h_2}u-\frac{h_2h_3}{h_1}y}~, \label{eq:6ptcond}
 \end{multline}
 where we have again used that the density of states factorizes at large energies. 
 In order to proceed, we will take the ansatz: 
 \begin{equation}
 \rho(h_1)\rho(h_2)\rho(h_3)\overline{\mathcal{C}_{OO O_1}\mathcal{C}_{OO O_2}\mathcal{C}_{OO O_3}\mathcal{C}_{O_1O_2O_3}}c(h_1,h_2,h_3)\underset{(h_1,h_2,h_3)\rightarrow\infty}{\sim} A_6^{-1}h_1^{\gamma_1-1}h_2^{\gamma_2-1}h_3^{\gamma_3-1}\label{eq:asymp6pt}
 \end{equation}
 and plug \eqref{eq:asymp6pt} into \eqref{eq:6ptcond}. This latter integral can be computed via saddle point and we find
 \begin{equation}
 	\gamma_1=\gamma_2=\gamma_3=2h_L~, \qquad\qquad A_6\approx 2^{6h_L-2}\pi^{3/2} e^{-6h_L+3}\left(h_L-\frac{1}{2}\right)^{6h_L-3/2}~. \label{eq:6ptgammaA}
 \end{equation}
 In this case $A_6$ is bounded from above by $\Gamma(2h_L)^3/4$.
 By comparing this observation with \eqref{eq:paramsintoverK} we  notice that
 \begin{equation}
 	A_6\approx 2 (A_4)^{3/2} \,,
 \end{equation}
 and that
 \begin{multline}
 	\rho(h_1)\rho(h_2)\rho(h_3)\overline{\mathcal{C}_{OO O_1}\mathcal{C}_{OO O_2}\mathcal{C}_{OO O_3}\mathcal{C}_{O_1O_2O_3}}c(h_1,h_2,h_3)\\\underset{(h_1,h_2,h_3)\rightarrow\infty}{\sim} A_6(h_L)^{-1}h_1^{2h_L-1}h_2^{2h_L-1}h_3^{2h_L-1}~. 
 \end{multline}
Of course, as was the case with the four-point function, we must combine this result with the anti-holomorphic dependence to obtain a statement about the asymptotic non-Gaussiantities surmised by this six-point analysis: 
  \begin{equation}
   \boxed{ 
   \begin{aligned}
   \rho(h_1)\rho(h_2)\rho(h_3)\rho(\bar h_1)\rho(\bar h_2)\rho(\bar h_3)&\overline{{C}_{OO O_1}{C}_{OO O_2}{C}_{OO O_3}{C}_{O_1O_2O_3}}c(h_1,h_2,h_3)c(\bar{h}_1,\bar{h}_2,\bar{h}_3)\\&\underset{(h_1,h_2,h_3,\bar h_1,\bar h_2,\bar h_3)\rightarrow\infty}{\sim}\ \frac{h_1^{2h_L-1}h_2^{2h_L-1}h_3^{2h_L-1}}{A_6(h_L)}\frac{\bar{h}_1^{2\bar{h}_L-1}\bar{h}_2^{2\bar{h}_L-1}\bar{h}_3^{2\bar{h}_L-1}}{A_6(\bar{h}_L)}
   \end{aligned}
   }~. \label{eq:asymp6ptboxed}
 \end{equation}
 \subsubsection{Gaussianity of the distribution?}
 An interesting test of the (non-)Gaussianity of the distribution over three-point coefficients is to check if we can estimate: 
 \begin{equation}
    \frac{\overline{{C}_{OO O_1}{C}_{OO O_2}{C}_{OO O_3}{C}_{O_1O_2O_3}}}{\sqrt{{C}^2_{OO O_1}{C}^2_{OO O_2}{C}^2_{OO O_3}}}~.
 \end{equation}
 Recall in section \ref{sec:sixpointallops}, this estimate fails because the right hand side depends on the dimension of $O$, $\Delta_L$ (see equation \eqref{ratioalld}). 
 However, for the case at hand, we find something different: 
\begin{multline}
	\frac{\overline{{C}_{OO O_1}{C}_{OO O_2}{C}_{OO O_3}{C}_{O_1O_2O_3}}}{\sqrt{{C}^2_{OO O_1}{C}^2_{OO O_2}{C}^2_{OO O_3}}}\\\underset{(h_1,h_2,h_3,\bar{h}_1,\bar{h}_2,\bar{h}_3)\rightarrow\infty}{\sim}\frac{\frac{ A_6(h_L)^{-1}h_1^{2h_L-1}h_2^{2h_L-1}h_3^{2h_L-1}}{\rho(h_1)\rho(h_2)\rho(h_3)c(h_1,h_2,h_3)}}{\sqrt{\frac{A_4(h_L)^{-1} h_1^{4h_L-1}}{\rho(h_1)c(h_1)}\frac{A_4(h_L)^{-1} h_2^{4h_L-1}}{\rho(h_2)c(h_2)}\frac{A_4(h_L)^{-1} h_3^{4h_L-1}}{\rho(h_3)c(h_3)}}}\times \text{anti. holo.}
\end{multline}
Plugging in the results from the previous sections, and using that the density of states is given by the Cardy formula, namely 
\begin{equation}
	\rho(h)=e^{2\pi\sqrt{\frac{c}{6}h}} \,,
\end{equation}
we find: 
\begin{multline}
	\frac{\overline{{C}_{OO O_1}{C}_{OO O_2}{C}_{OO O_3}{C}_{O_1O_2O_3}}}{\sqrt{{C}^2_{OO O_1}{C}^2_{OO O_2}{C}^2_{OO O_3}}}{\sim} \\ \frac{\sqrt{c(h_1)c(h_2)c(h_3)c(\bar h_1)c(\bar h_2)c(\bar h_3)}}{c(h_1,h_2,h_3)c(\bar h_1,\bar h_2,\bar h_3)}\times \frac{e^{-\pi\sqrt{\frac{c}{6}}\left(\sqrt{h_1}+\sqrt{h_2}+\sqrt{h_3}\right)}}{2\sqrt{h_1\,h_2\,h_3}} \times \frac{e^{-\pi\sqrt{\frac{c}{6}}\left(\sqrt{\bar h_1}+\sqrt{\bar h_2}+\sqrt{\bar h_3}\right)}}{2\sqrt{\bar h_1\,\bar h_2\,\bar h_3}}~.  
\end{multline}
The dependence on the light data has completely dropped out! Perhaps this indicates a type of Wick's theorem for the asymptotics of higher-point non-Gaussiantities in the case of quasi-primary operators. For Virasoro primaries, as we will now see, the Wick rule no longer applies.

\section{The Virasoro Crossing Kernel and \texorpdfstring{$k$}{k}-point Crossing  \label{sec4}}

In this section, we present a method for deriving the OPE statistics of Virasoro primary operators in $d=2$, obtained from the crossing kernel. The setting is that of irrational, compact and unitary CFTs. More precisely, 2d CFTs with $c>1$, a discrete spectrum, an infinite number of primary states with a unique $\mathfrak{sl}(2)$-invariant ground state, and in the absence of any extended chiral algebra. 

As in previous sections we will derive an asymptotic constraint on OPE data obtained by considering crossing symmetry of the six-point function in the star channel. But it turns out that we can generalize this result to higher moments of OPE coefficients by considering crossing of the $k$-point function. 

In this section, we will work with \emph{Liouville} variables $b, Q, P$ and $\alpha$, natural for analyzing the Virasoro crossing kernel,  defined as
\begin{gather}
    c = 1+6 Q^2 = 1+6\big(b+b^{-1}\big)^2,\\
    h = P^2 + \frac{Q^2}{4}= \alpha (Q-\alpha), \qquad \alpha = \frac{Q}{2}+iP.
\end{gather}
We will cover the $c>25$ branch with $b<1$. For $1<c<25$, $b$ will be restricted to a pure phase in the first quadrant. 

The definition of $h$ is invariant under $P\rightarrow -P$ and/or $\alpha \rightarrow Q-\alpha$, and there are two different noteworthy ranges of $h$: the discrete range encompassing $0 \leq h<(c-1)/24$, and the continuum, where $h\geq (c-1)/24$. For $h$ in the discrete range, our conventions will be: $\alpha \in [0,Q/2)$ and $P \in i (0, Q/2]$. In the continuum range, we will take $P\in \mathbb{R}$ and $\alpha \in Q/2 + i \mathbb{R}$. This is summarized here for ease of reading: 
\begin{align*}
&\alpha \in [0,Q/2)~, ~P \in i (0, Q/2] & \text{discrete range:} ~~0 \leq h<(c-1)/24~,\\
&P\in \mathbb{R}~, ~\alpha \in Q/2 + i \mathbb{R} & \text{continuum range:} ~~ h\geq (c-1)/24~.
\end{align*}
We will refer to \emph{both} $\alpha$ and $P$ as \emph{momenta}. Unless stated otherwise, we will work with $P$ in the continuum and $\alpha$ in the discrete range. 

\subsection{The six-point function}

\begin{figure}
    \centering
    \includegraphics{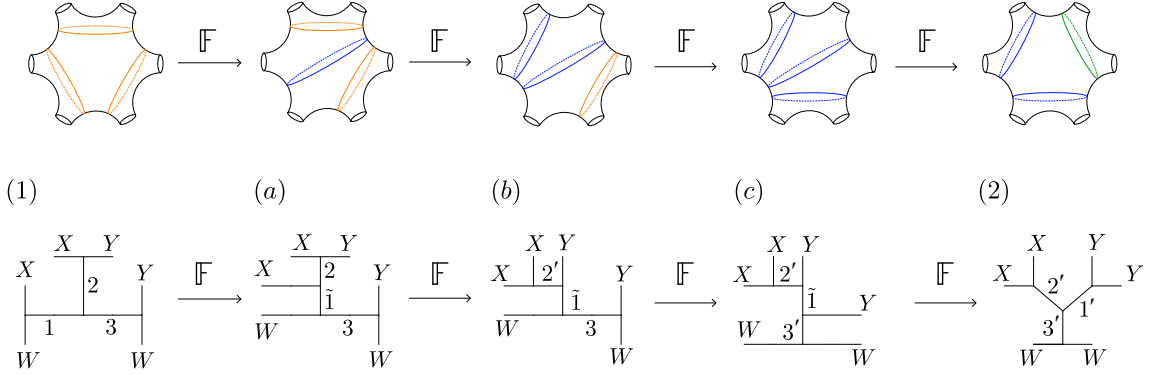}
    \caption{The sequence of elementary crossing moves relating two different star channel decompositions. We show the full (top) and simplified (bottom) versions of this diagram.}
    \label{six-pointkernel}
\end{figure}

The previous section considered the six-point function of six quasi-primary operators $O$ decomposed in \emph{global} conformal blocks. This analysis allowed us to extract results for the averaged non-Gaussianities of OPE coefficients involving quasi-primary states. In two dimensions, however, the Virasoro algebra allows us to decompose our correlation functions in terms of \emph{Virasoro}-primaries $\mathcal{O}$, and the OPE coefficients involving Virasoro primaries can and will have different statistics. 

One of the main obstacles when generalizing the global conformal analysis to the Virasoro case is that closed-form expressions for the Virasoro conformal blocks are only known in exceptional cases. The most effective way to compute Virasoro blocks is numerically \cite{Zamolodchikov:1984eqp,zamo,Hadasz:2009db,Cho:2017oxl}, but even then, it remains challenging to formulate crossing symmetry using them. Moreover, since the Virasoro conformal blocks do not satisfy simple differential equations like their global counterparts, the analysis of the previous section can not be generalized. Fortunately, there is another method -- based on the crossing kernel -- that allows us to extract OPE statistics \cite{Collier:2019weq}, and remarkably, this method does not require prior knowledge of the blocks.\footnote{See \cite{Kusuki:2019gjs} for a previous discussion of this method for higher-point correlation functions.}

We start by writing the six-point function of Virasoro primaries in two different star channel decompositions (see the rightmost and leftmost diagrams in Figure \ref{six-pointkernel})
\begin{align}
\nonumber
    \langle\clw(w_2,\bar w_2) \clx(x_1,&\bar  x_1)\clx(x_2, \bar x_2)\cly(y_1,\bar y_1)\cly(y_2,\bar y_2)\clw(w_1,\bar w_1) \rangle
    \\ \label{channels1} & =  \int \left( \prod_{i=1}^3 \frac{dP_i d\bar P_i}{4}\den(P_i)\den(\bar P_i)\right)\vd_{WXY}^{(1)}(P_1,P_2,P_3,\bar P_1,\bar P_2,\bar P_3)\times \clf^{(1)} \bar \clf^{(1)}
    \\ \label{channels2} & =  \int \left( \prod_{i=1}^3 \frac{dP_i' d\bar P_i'}{4}\den(P_i')\den(\bar P_i')\right)  \vd_{WXY}^{(2)}(P_1',P_2',P_3',\bar P_1',\bar P_2',\bar P_3')\times \clf^{(2)} \bar \clf^{(2)}.
\end{align}
In this expression, $\clf^{(i)} = \clf^{(i)}(P_\clw,P_\clx,P_\cly; P_1,P_2,P_3; x_1\cdots w_2)$ denotes the Virasoro conformal block associated to the six-point OPE densities
\begin{align}
    \vd_{WXY}^{(1)}(P_i,\bar P_i) &= \sum_{\clo_1\clo_2\clo_3}  \frac{\C_{WX\clo_1}\C_{XY\clo_2}\C_{YW\clo_3}\C_{\clo_1\clo_2\clo_3}}{\prod_{l=1}^3\den(P_l)\den(\bar P_l)} \prod_{l=1}^{3}\delta_s(P_l,P_{\clo_l})\delta_s(\bar P_l,\bar P_{\clo_l})\label{Qexpression1} \\
    \vd_{WXY}^{(2)}(P_i',\bar P_i') &= \sum_{\clo_1'\clo_2'\clo_3'}  \frac{\C_{XX\clo_2'}\C_{YY\clo_1'}\C_{WW\clo_3'}\C_{\clo_1'\clo_2'\clo_3'}}{\prod_{l=1}^3\den(P_l')\den(\bar P_l')} \prod_{l=1}^{3}\delta_s(P_l',P_{\clo_l'})\delta_s(\bar P_l',\bar P_{\clo_l'}),
\end{align}
where $\den(P_i)$ denotes the density of primary states, which at large (positive) $P_i$ is given by Cardy's formula $\den(P_i) \sim \exp{2\pi Q P_i}$, and the sums are over the primary fields $\clo_i$ of the theory. We are also using the shorthand notation $\delta_s(x,y) = \delta(x+y)+\delta(x-y)$; our measures and densities are defined in such a way that they take into account the  $P_i\rightarrow - P_i$ redundancy in the momenta. 

The crossing kernel $\bbk$ that relates the two conformal block decompositions is defined via the equation 
\begin{multline}
\label{Vir1}
    \clf^{(2)}(P_W,\cdots;P_1',P_2',P_3';x_1\cdots) \\= \int \left( \prod_{i=1}^3 \frac{dP_i}{2}\right) \times \bbk_{\{P_1,P_2,P_3\},\{P_1',P_2',P_3'\}}\clf^{(1)}(P_W,\cdots;P_1,P_2,P_3;x_1\cdots).
\end{multline}
For correlation functions on the sphere, these kernels always exist and in fact can be built from an elementary building block: the \emph{Virasoro fusion kernel}, F-transform or 6$j$ symbol $\bbf$. The process of building general crossing kernels using elementary crossing ``moves" goes under the name of the Moore-Seiberg construction \cite{Moore:1988uz,Moore:1988qv}. This process is not limited to the study of correlation functions on the sphere, but it can also be applied to more general observables on different Riemann surfaces. We recommend  \cite{Belin:2021ryy,Kusuki:2018wpa,QuantumRegge,Collier:2019weq} for a detailed discussion of the topic. An important property about these kernels is that they are kinematic objects -- i.e. they depend only on the central charge of the theory and the conformal dimension of the operators involved but not on the OPE data. This is clear from the definition, which only involves Virasoro conformal blocks.

We can build the crossing kernel relating $\clf^{(1)}$ and $\clf^{(2)}$ in stages. For concreteness, let us denote the intermediate conformal blocks described in Figure \ref{six-pointkernel} by $\clf^{(a)},\clf^{(b)},\clf^{(c)}$. The first move in the crossing kernels reads
\begin{equation}
\clf^{(2)}(P_W,\cdots,P_1',P_2',P_3',x_1\cdots) = \int \frac{d\tilde P_1}{2} 
\fker{\tilde P_1}{P_1'}{P_3'}{P_Y}{P_Y}{P_2'} \clf^{(c)}(P_W,\cdots,\tilde P_1,P_2',P_3',x_1\cdots) \,.
\end{equation} 
Note that in this transformation (as in the definition of the crossing kernel) the measure does not include the usual factor of $\den$. This is just a convention when defining the fusion kernel. The rest of the crossing moves, $\clf^{(c)} = \bbf \clf^{(b)}$, $\clf^{(b)} = \bbf  \clf^{(a)}$, etc., look schematically the same. Connecting these equations, we find the crossing kernel relating $\clf^{(1)}$ and $\clf^{(2)}$: 
\begin{equation}
\bbk_{\{P_1,P_2,P_3\},\{P_1',P_2',P_3'\}} = \int\frac{d\tilde P_1}{2}\; \fker{P_1}{\tilde P_1}{P_3}{P_2}{P_\clx}{P_\clw}
\fker{P_2}{P_2'}{P_Y}{P_X}{P_X}{\tilde P_{1}}
\fker{P_3}{P_3'}{P_W}{P_Y}{\tilde P_1}{P_W}
\fker{\tilde P_1}{ P_{1}'}{P_3'}{P_Y}{P_Y}{P_2'}~.\label{eq:vircrosskernel}
\end{equation}
In \rref{Vir1}, we are only integrating over the unprimed variables $P_i$, so we have to interpret the integral over $\tilde P_1$ as being part of the crossing kernel. Naturally, the crossing kernel only depends on the $P_i$ and $P'_i$ momenta.

We give a closed expression for the fusion kernel $\bbf$ and a summary of its properties in appendix \ref{appendixfusion}. Plugging \rref{Vir1} into \rref{channels2} and comparing the result with \rref{channels1} gives us the crossing equation:
\begin{equation}
\label{crossEq}
\begin{split}
   & \vd_{WXY}^{(1)}(P_i,\bar P_i) \\&= \frac{1}{\prod_{l=1}^{3}\den(P_l)\den(\bar P_l)} \int  \left( \prod_{l=1}^3 \frac{dP_l' d\bar P_l'}{4}\den(P_l')\den(\bar P_l')\right)\\ &\eqspace{19em} \times\bbk_{\{P_i\},\{P_i'\}}\bbk_{\{\bar P_i\},\{\bar P_i'\}}  \vd^{(2)}_{WXY}(P_i',\bar P_i')\\& =
    \frac{1}{\prod_{l=1}^{3}\den(P_l)\den(\bar P_l)} \sum_{\clo_1'\clo_2'\clo_3'}\!\! \C_{XX\clo_2'}\C_{YY\clo_1'}\C_{WW\clo_3'}\C_{\clo_1'\clo_2'\clo_3'} \bbk_{\{P_{i}\},\{P_{\clo'_i}\}}\bbk_{\{\bar P_i\},\{\bar P_{\clo'_i}\}}.
\end{split}
\end{equation}
In the second line, we have used the definition of $\vd^{(2)}$ in terms of delta functions. Informally, we may think of the crossing kernel as the \emph{change-of-basis} matrix relating the two conformal block decompositions $\clf^{(1)}$ and $\clf^{(2)}$. In this setting, the conformal blocks form a basis of the space of correlation functions and the OPE coefficients correspond to the vector components of the six-point function. 

We now have two expressions for $\vd_{WXY}^{(1)}$ given by \eqref{Qexpression1} and \eqref{crossEq}. If we can somehow project the $\mathcal{O}_i'$ in \eqref{crossEq} to the identity, using a feature of the crossing kernel $\bbk$, we will be able to derive an asymptotic formula. This is achieved by taking large values of $P_i$ and $\bar P_i$. To realize this projection, we will focus on the following heavy limit: 
\begin{equation}
\label{heavylimit}
  P_i= P+ \delta_i ~, \qquad\qquad \text{and}\qquad\qquad |P| \gg c, |\delta_i|~,
\end{equation}
 for $i=1,2,3$.
In this heavy limit, we will use \eqref{Qexpression1} to write: 
\begin{equation}
   \lim_{|P|\rightarrow\infty}  \vd_{WXY}^{(1)} =\overline{\C_{WX\clo_1}\C_{XY\clo_2}\C_{YW\clo_3}\C_{\clo_1\clo_2\clo_3}}
\end{equation}
as we did in equation \eqref{eq:globallargeHlimit} for quasi-primaries. Furthermore, analyzing \eqref{crossEq} reveals that this alternate expansion is dominated by the exchange of the identity operator: 
\begin{equation}
\label{denEqu}
    \vd^{(1)}_{WXY} =
   \lim_{|P|\rightarrow\infty} \frac{\bbk_{\{P_i\},\{\bbi\}}\bbk_{\{\bar P_i\},\{\bbi\}}}{\prod_{l=1}^{3}\den(P_l)\den(\bar P_l)}\left[1+\ord\Big(e^{-2\pi\alpha_\chi |P|}e^{-2\pi\bar \alpha_\chi |\bar P|}\Big)\right],
\end{equation}
where $\alpha_\chi$ corresponds to the momentum of the second lightest primary operator in the theory that contributes to the sum in \rref{crossEq}. The disappearance of the three-point coefficients stems from the fact that $C_{XX\bbi}=1$ and so on, since the operators are normalized to have unit norm. The result \rref{denEqu} follows because 
\begin{equation}
\label{technicalsixpoint}
    \frac{\bbk_{\{P_1,P_2,P_3\},\{P_1',P_2',P_3'\}}}{\bbk_{\{P_1,P_2,P_3\},\{\bbi,\bbi,\bbi\}}} \sim e^{-2\pi(\alpha'_2 + \alpha_3')|P|},
\end{equation}
when $|P|\rightarrow\infty$, which we derive in appendix \ref{appendixSixpoint}. Technically, the crossing kernel \eqref{eq:vircrosskernel}, at large $|P|$ does not project to the identity for $\clo_1'$. However, by virtue of the OPE coefficient $C_{\clo_1'\clo_2'\clo_3'}$, it is sufficient to project $\clo_2'$ and $\clo_3'$ to the identity in order to subsequently project $\clo_1'$ to the identity in tandem. 

 Since the leading order term in this equation is a crossing kernel, we have derived a universal result for this product of averaged OPE coefficients. An evaluation of the crossing kernel in this limit, see \rref{crossKer}, yields the result 
\begin{equation}
\label{sixpointstat}
    \overline{\C_{WX\clo_1}\C_{XY\clo_2}\C_{YW\clo_3}\C_{\clo_1\clo_2\clo_3}} \sim \left(\frac{3\sqrt{3}}{16}\right)^{3P^2}e^{-\frac{9 \pi}{2} Q |P|}|P|^{-\frac{5+11Q^2}{6}+4(h_W+h_X+h_Y)}\times \textnormal{(a.c.)} \,,
\end{equation}
where (a.c.) stands for the anti-holomorphic counterpart of this expression. Note that this expression drops terms proportional to $\delta_i-\delta_j$ as these terms are not necessarily illuminating and are subleading in the large $|P|$ limit. As a function of scaling dimension $\Delta$, we have

\bea \label{result1sec4}
 \overline{\C_{WX\clo_1}\C_{XY\clo_2}\C_{YW\clo_3}\C_{\clo_1\clo_2\clo_3}} \sim \left(\frac{3\sqrt{3}}{16}\right)^{3\Delta_H}e^{-\frac{9 \pi}{2} \sqrt{
\frac{c-1}{3}\Delta_H}}\Delta_H^{6\Delta_L-\frac{19+11c}{36}}  \,.
\eea

\subsection{The $k$-point function}

In the following, we will present a slightly different perspective on how to use these kernels. Our goal is to show that we can use them to simplify more general OPE densities. This will allow us to write down a result for the OPE statistics of the $k$-point function. 

The starting point of our analysis is again the six-point function in \eqref{Qexpression1}. Instead of considering the entire crossing kernel, let us examine the kernel relating $\clf^{(1)}$ and $\clf^{(b)}$ (see Figure \ref{six-pointkernel} where the superscript labels the step represented in the figure):
\begin{multline}
\clf^{(b)}(P_W,\cdots;\tilde P_1,P_2',P_3,x_1,\cdots) = \int \frac{dP_1}{2}\frac{dP_2}{2} 
\fker{P_1}{\tilde P_1}{P_3}{P_2}{P_X}{P_W}\fker{P_2}{P_2'}{P_Y}{P_X}{P_X}{\tilde P_1} \\ \times \clf^{(1)}(P_W,\cdots;P_1,P_2,P_3,x_1,\cdots),
\end{multline}
that leads to the crossing equation 
\begin{equation}
\label{Vir2}
\begin{split}
\vd_{WXY}^{(1)}(P_1,P_2,P_3) &= \frac{1}{\den(P_1)\den(P_2)} \int \frac{d \tilde P_1}{2}\frac{dP_2'}{2} \den(\tilde P_1)\den(P_2') \fker{P_1}{\tilde P_1}{P_3}{P_2}{P_X}{P_W}\fker{P_2}{P_2'}{P_Y}{P_X}{P_X}{\tilde P_1}\\ & \eqspace{23em} \times \vd^{(b)}_{WXY}(\tilde P_1,P_2',P_3)\\
&=
\frac{1}{\den(P_1)\den(P_2)}\sum_{\tilde \clo_1 \clo_2'\clo_3} \frac{C_{XX\clo_2'}C_{\clo_2'Y\tilde \clo_1} C_{W\tilde \clo_1\clo_3} C_{\clo_3 YW}}{\den(P_3)}\\ & \eqspace{13.2em} \times 
\fker{P_1}{P_{\tilde \clo_1}}{P_3}{P_2}{P_X}{P_W}\fker{P_2}{P_{\clo_2'}}{P_Y}{P_X}{P_X}{P_{\tilde \clo_1}}\delta_s(P_3,P_{\clo_3}).
\end{split}
\end{equation}
In this expression, and in the equations that follow, we omit the anti-holomorphic parts for greater clarity. In the second line of this equation, we have replaced the density by its definition 
\begin{equation}
 \vd_{WXY}^{(b)}(\tilde P_1,P_2',P_3) = \sum_{\tilde \clo_1\clo_2'\clo_3}  \frac{C_{XX\clo_2'}C_{\clo_2'Y\tilde \clo_1}C_{W\tilde \clo_1\clo_3}C_{\clo_3YW}}{\den(\tilde P_1)\den(P_2')\den(P_3)}  \prod_{l=\tilde 1,2',3}\delta_s(P_l,P_{\clo_l}).
\end{equation}

\begin{figure}[t]
\centering
    \includegraphics[trim={0 4cm 0 0},clip]{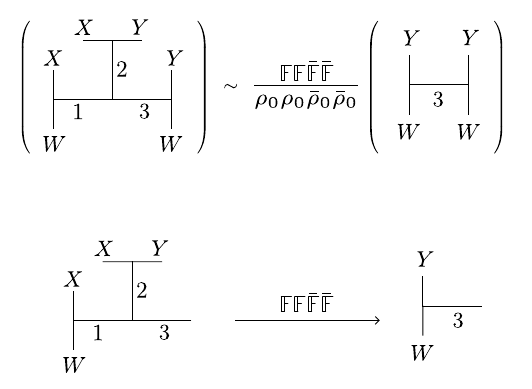}
    \caption{The six-point OPE density associated with the channel on the left can be approximated by the four-point density on the right multiplied by a crossing kernel.}
    \label{Relation}
\end{figure}

Now, we would like to simplify equation \eqref{Vir2} by projecting the sum over $\clo_2'$ to the identity. This is analogous to what we did in the previous section. We need to project the sum over $\clo_2'$ and not over $\tilde \clo_1$ because the OPE coefficient $C_{W\tilde \clo_1\clo_3}$ would otherwise vanish; note that the identity exchange in $\clo_2'$ fixes the exchange over $\tilde \clo_1$ via the coefficient $C_{\clo_{2}'Y\tilde \clo_1}$. In the heavy limit \eqref{heavylimit}, we find that the sum over $\clo_2'$ is exponentially suppressed by the term $\exp\!\{-2\pi \alpha_{\clo_2'}|P|\}$. This can be seen from the asymptotic expansion of the kernels
\begin{multline}
\fker{P_1}{P_{\tilde \clo_1}}{P_3}{P_2}{P_X}{P_W}\fker{P_2}{P_{\clo_2'}}{P_Y}{P_X}{P_X}{P_{\tilde \clo_1}} \sim \left(\frac{81\sqrt{3}}{256}\right)^{P^2}e^{
\big[\frac{\pi}{2}(Q-4\alpha_{\clo_2'}) -(\delta_1+\delta_2)\log\frac{256}{27}+\delta_3\log 27\big]|P|}\\\ \times |P|^{-\frac{Q^2+1}{3}+4h_{X}+2(h_{Y}-h_{\tilde \clo_1})}.
\end{multline}
Since the identity exchange dominates at large $|P|$, we have
\begin{multline}
\vd^{(1)}_{WXY}(P_1,P_2,P_3) =\lim_{|P|\rightarrow\infty}
\frac{\fker{P_1}{P_Y}{P_3}{P_2}{P_X}{P_W}\fker{P_2}{\bbi}{P_Y}{P_X}{P_X}{P_{Y}}}{\den(P_1)\den(P_2)}
\\\times\left( \sum_{\clo_3} \frac{C_{WY\clo_3}^2}{\den(P_3)}\delta_s(P_3,P_{\clo_3})\right)\Big[1 + \ord\left(e^{-2\pi \alpha_\chi |P|}\right)\Big].
\end{multline}
This equation describes a relationship between two different OPE densities. Inside the parenthesis, we have the density of the four-point function $\langle WYYW \rangle$ in the $(WY)(YW)$ channel (see Figure \ref{Relation}). This density has been studied extensively and, using crossing kernel arguments, one can show that \cite{QuantumRegge, Collier:2019weq, Belin:2021ryy,Das:2017cnv}
\begin{equation}
\left( \sum_{\clo_3} \frac{C_{WY\clo_3}^2}{\den(P_3)}\delta_s(P_3,P_{\clo_3})\right) = \lim_{|P|\rightarrow\infty}\frac{ \fker{P_3}{\bbi}{P_Y}{P_W}{P_W}{P_Y}}{\den(P_3)}\Big[1 + \ord\left(e^{-2\pi \alpha_\chi |P|}\right)\Big].
\end{equation}       
After substituting the different asymptotic formulas for the crossing kernels, we recover the result of the previous section\footnote{In the final result, we omitted the factor $-(\delta_1+\delta_2+\delta_3)P\log\frac{256}{27}$ because it is subleading.} \eqref{sixpointstat}.

\begin{figure}[t]
    \centering
    \includegraphics[trim={0 0 0 4cm},clip]{recursion.pdf}
    \caption{The OPE diagram on the left can be reduced to the one on the right by an application of two crossing moves.}
    \label{k-pointRecursion}
\end{figure}

\begin{figure}[t]
    \centering
    \includegraphics{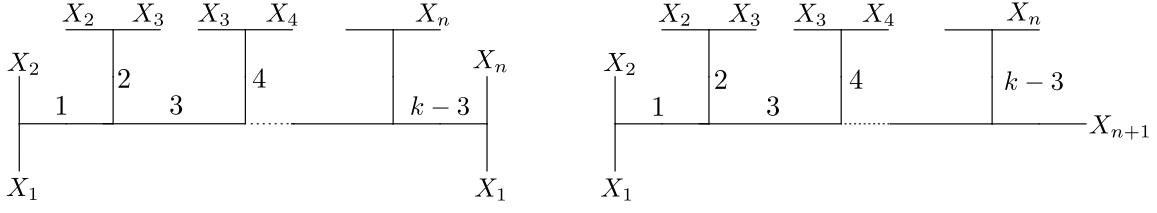}
    \caption{The trivalent diagrams associated to the OPE density $\vd_k$. These diagrams correspond to one of the possible OPE channel decompositions of the  $k$-point function for even (\emph{left}) $k =2n$ or odd (\emph{right}) $k=2n-1$ values of $k$.}
    \label{defGk}
\end{figure}

The important part of this construction is that it holds more generally as the crossing kernel acts ``locally" on a diagram of OPE coefficients. If we have a configuration like the one depicted in Figure \ref{k-pointRecursion}, we can always use the same crossing moves to remove two of the heavy lines and the external operators in the middle. This simplification is valid as long as we are working in the heavy limit with fixed differences. For the eight-point function in the star channel, for instance, this implies that
\begin{multline}
\vd_8(P_{X_1},P_{X_2},P_{X_3},P_{X_4}; P_1, P_2,P_3,P_4,P_5) \approx \\
\frac{
\fker{P_1}{P_{X_3}}{P_3}{P_2}{P_{X_2}}{P_{X_1}}
\fker{P_2}{\bbi}{P_{X_3}}{P_{X_2}}{P_{X_2}}{P_{X_3}}}{
\den(P_1)\den(P_2)}
\vd_{6}(P_{X_1},P_{X_3},P_{X_4};P_3,P_4,P_5).
\end{multline}
After inserting the formulas corresponding to the kernels and the six-point density, this yields an asymptotic expression for the eight-point function. The densities $Q_8$ and $Q_6$ are defined in Figure \ref{defGk} for the case of the $k$-point function. To find an expression for $Q_k$, where $k$ is even, we just need to repeat this process $(k-4)/2$ times; reducing $Q_k$ to the four-point function OPE density. The final result is that  
\begin{equation} \label{resultVireven}
    \vd_k \sim 3^{\frac{9(k-4)}{4}\Delta}16^{(3-k)\Delta}e^{\frac{24-7k}{4}\pi Q (P+\bar P)}\left(P\bar P\right)^{-\frac{(k-1)+(5+k)Q^2}{6}}P^{\sum_{i=1}^{k/2}4 h_{X_i}}\bar P^{\sum_{i=1}^{k/2} 4\bar h_{X_i}}.
\end{equation}
Here, we have restored the anti-holomorphic contributions of the OPE density. 

To find an expression for $Q_k$ when $k$ is odd, we need to know the statistics of an odd diagram, like $Q_5$. Then, we can apply our crossing moves to find asymptotic formulas for $Q_7$, $Q_9$, etc. We relegate the details of this computation to appendix \ref{appendixFivepoint}, where we find that 
\begin{equation}
Q_5(P_{X_1},P_{X_2},P_{X_3},P_{X_4};P_1,P_2) \approx \frac{C_{X_3X_1X_4}}{\den(P_1)\den(P_2)}\fker{P_1}{P_{X_3}}{P_{X_4}}{P_2}{P_{X_2}}{P_{X_1}}
    \fker{P_2}{\bbi}{P_{X_3}}{P_{X_2}}{P_{X_2}}{P_{X_3}}.
\end{equation}
Reducing $Q_k$ to $Q_5$, requires us to apply $(k-5)/2$ times the sequence of moves depicted in Figure \ref{k-pointRecursion}. After which, we find that 
\begin{align} \label{resultVirodd}
\nonumber 
 \vd_k \sim & C_{X_1X_{n}X_{n+1}}3^{\frac{9(k-5)}{4}\Delta}16^{(4-k)\Delta} e^{\frac{23-7k}{4}\pi Q (P+\bar P)}\left(P\bar P\right)^{-\frac{(k-2)+(k+4)Q^2}{6}}\\ \nonumber &\times P^{2h_{X_{n-1}}-2h_{X_n}-3h_{X_{n+1}}-(P_{k-3}-P_{k-4})^2+\sum_{i=2}^{n+1}4h_{X_i}} \\ &\times \bar P^{2\bar h_{X_{n-1}}-2\bar h_{X_n}-3\bar h_{X_{n+1}}-(\bar P_{k-3}-\bar P_{k-4})^2+\sum_{i=2}^{n+1}4\bar h_{X_i}}.
\end{align}
The expression for both odd and even $k$ as a function of the scaling dimension $\Delta$ is given in \rref{resultallpoint}.

Interestingly, for even $k$ the result is a universal formula, while for odd $k$ the result explicitly depends on the OPE coefficient $C_{X_1X_nX_{n+1}}$. In both cases, we see that the addition of two more operators to the correlation function corresponds to an additional suppression of the OPE by $-7/4$ factors of the entropy $S=2\pi Q(P+\bar P)$. This is the expected behaviour for the distribution of OPE coefficients, with higher moments being further exponentially suppressed.

There is an important caveat worth emphasizing for the odd-point case. The formula \rref{resultVirodd} only holds if $C_{X_1X_nX_{n+1}}\neq0$. For identical external operators, this requires $C_{OOO}\neq0$ which could easily be violated, for example, if $O$ has a $\mathbb{Z}_{2}$ symmetry. Interestingly, we were not able to find an asymptotic formula if $C_{X_1X_nX_{n+1}}=0$. The reason is the following. Consider the five-point function, decomposed as in Figure \ref{fivepointrecursion}. For the order in which we applied the crossing transformations, the sum over operators $2'$ is exponentially suppressed. This sets $2'=\bbi$, which in turn sets $1'=X_3$. Then we are left with an OPE coefficient $C_{X_1X_3X_4}$ which we need to assume does not vanish. Note that the sum over $1'$ is not exponentially suppressed, but there is a selection rule for $1'$ as long as $2'=\bbi$. If $C_{X_1X_3X_4}=0$, then we must go to the lightest non-trivial operator that appears in the $2\times2$ OPE (let us call it $O'$). But we are now faced with a problem: the sum over $1'$ is not exponentially suppressed, and we must sum over all operators $1'$ such that $C_{O'1'X_3}\neq0$. There is no simple way to perform this sum which prevents us from obtaining an asymptotic formula. Similar effects were encountered at higher genus in \cite{Belin:2021ryy}. Both for the higher genus case and for the present result, it would be interesting to understand the origin of this obstruction from a gravity point of view.

\section{Discussion  \label{sec5}}

This paper is concerned with deriving new asymptotic formulas for the products of OPE coefficients in conformal field theory. This is achieved using crossing symmetry of the six- and higher-point functions in the Euclidean OPE limit, and results in formulas involving combinations of light-light-heavy and heavy-heavy-heavy OPE coefficients. We presented expressions valid in arbitrary dimensions for the asymptotics of OPE coefficients involving all operators of the theory (both primary and descendants) and also derived refinements in the two-dimensional case. These $d=2$ results include asymptotic formulas for products of OPE coefficients of quasi-primaries or Virasoro primaries. Because our formulas involve at least four OPE coefficients, they should be interpreted as probing non-Gaussianities of the distribution of OPE coefficients. We conclude here with some open questions and future directions.

\subsection{6pt blocks and primaries vs descendants}

For CFTs in $d>2$ dimensions, we  present asymptotic formulas  for operators, irrespective of their status as primary or descendant fields. This is  coarse information, as representation theory of the conformal group can, in principle, be used to separate the contribution of primary and descendant operators. For the four-point function, this was achieved in \cite{Mukhametzhanov:2018zja}. To obtain  higher-point statistics involving primary fields, one would need to either have good control over six-point conformal blocks, or to have a closed-form expression for the crossing kernel at six points. There has been some progress in constructing six-point conformal blocks in $d>2$ \cite{Fortin:2019zkm,Fortin:2020yjz,Buric:2021ttm,Buric:2021ywo}, but they are represented as infinite sums of hypergeometric functions and thus it is difficult to extract bounds on these functions' asymptotic behavior in the cross-channel.  One interesting avenue is to make use of  the explicit expression for the 6$j$ symbols given in \cite{Liu:2018jhs} towards the study of six-point crossing. 

Overall, our analysis clearly shows that the asymptotic formulas are drastically different depending on whether they count primaries or all operators. The contribution of descendants is therefore not negligible, even if most heavy operators are primaries. This fact had already been observed for the lower moments, as early as  \cite{Pappadopulo:2012jk}. It would be interesting to develop a more intuitive way to understand these differences.

\subsection{Size of the coarse-graining window}

We would like to emphasize that the asymptotic formulas derived here and elsewhere do not imply ETH or the ORH. This is because they are obtained from crossing symmetry (or modular invariance) and therefore hold in both chaotic and integrable theories. To wit, the formulas derived from crossing symmetry on the plane are even satisfied by mean field theory (a.k.a. generalized free theories), in stark contrast to where ETH should hold. Asymptotic formulas for chaotic versus integrable CFTs may nevertheless be distinguished by the size of the averaging window one needs in order to obtain a good estimate for the OPE density. In particular, the size of this window is expected to be parametrically smaller in chaotic theories that have a very finely spaced spectrum. 

Tauberian theory \cite{Qiao:2017xif,Mukhametzhanov:2018zja,Pal:2019zzr,Mukhametzhanov:2019pzy,Mukhametzhanov:2020swe,Das:2020uax} allows us to obtain bounds on these window sizes. In particular, one may try to find an upper bound on the size of these averaging windows which holds for arbitrary CFTs.\footnote{Note that there is always an estimate for the window size that one can extract from the consistency of the saddle-point analysis, but that estimate scales with some positive power of the energy. This is a gross overestimate, even in free theories.} From the point of view of ETH, where an eigenstate is representative of the entire ensemble, one is prompted to ask: Under what conditions does this window size become parametrically small? For example, one could take the window to be of size $\frac{M}{\rho(\Delta)}$ for $M\gg 1$ but independent of the density of states. For an integrable theory like our mean field theory example, this window would generically contain no operators at all, and hence not be controlled by the asymptotic formula. However, for a chaotic theory, it would contain on average $M$ operators and thus give a good estimate of the OPE density for large $M$. 
To the best of our knowledge, this remains an important open problem which deserves further investigation.

\subsection{Are higher moments suppressed?}

Having underlined that asymptotic formulas cannot provide a direct justification for the ORH, one can nevertheless ask what their implications are if the ORH were  satisfied.  Note that, for the ORH to hold, it is crucial that higher moments of the distribution over OPE coefficients be further suppressed in the entropy $S$ compared to lower moments \cite{Foini:2018sdb}. For coefficients involving any operator in $d>2$ (or for quasi-primaries in $d=2$), the entropic suppression of the higher moments seems to be  universal. For example, the Gaussian part of the distribution gives\footnote{Note that the terms like $\Delta_H^{2\Delta_L-1}$ are completely irrelevant here, as they only give power-law correction in the entropy.}
\be
\overline{C_{LLH}^2} \sim e^{-S} \,.
\ee
From our analysis involving crossing symmetry of the six-point function, we found:
\be
\overline{C_{LLH}^3C_{HHH} }\sim e^{-3S} \,.
\ee
However, since this formula involves OPE coefficients of LLH- and HHH-type, it is not immediately obvious how one should ascribe entropic suppressions between the different elements of the above formula, since we have no independent calculation of $C_{HHH}$. If we imagine distributing the entropic factors such that we have $C_{HHH}\sim e^{-\beta S}$, we would find
\be \label{discussionCLLH}
C_{LLH}|_{\textrm{2nd moment}} \sim e^{-S/2} \,, \qquad C_{LLH}|_{\textrm{4th mixed moment}} \sim e^{-(3-\beta)S/3} \,.
\ee
This would mean that the higher moments of $C_{LLH}$ are suppressed compared to the lower moments, if $\beta<3/2$. It is worth mentioning that a similar reasoning will continue for the higher moments, even though we have not computed them in this paper. This follows from a simple counting argument for the number of powers of the density of states that can appear in any given correlation function. Because the poles we are trying to reproduce have an $\mathcal{O}(1)$ coefficient in the entropy and since the blocks do not give any exponential factors in the entropy, we can simply  count the number of operators that are summed over in the asymptotic formula. This guarantees further exponential suppression of the higher moments.

In $d=2$ for Virasoro primaries, the situation is somewhat different. The crossing kernel provides us with expressions for the arbitrary-point moments of OPE coefficients as presented in  \rref{resultVireven} and \rref{resultVirodd}. We again see that the higher moments come with an exponential suppressions in the entropy, but this time the Virasoro  blocks are absorbing some factors of $e^S$. Unlike the moments of global OPE coefficients presented above, we have not managed to find an intuitive explanation for the integers appearing in front of $S$ in the exponential of  \rref{resultVireven}. Precise features of the crossing kernel are responsible for the entropic suppression that appears, for example the second moment is found to be:  
\be
C_{LLH}|_{\textrm{2nd moment}}^{\textrm{Vir. prim.}} \sim e^{-S/4} \,,
\ee
which should be contrasted with the first equation of \rref{discussionCLLH}. One can also compute the exponential suppression of the distribution over the six-point function with respect to its Gaussian counterpart. The quantity to compute is
\begin{equation}
\frac{\overline{C_{LLH}^3C_{HHH}}}{\sqrt{\left(\overline{C_{LLH}^2}\right)^3\overline{C_{HHH}^2}}}  \,.
\end{equation}
To proceed, we need the asymptotic formula derived in this paper, as well as that of $\overline{C_{LLH}^2}$ which comes from four-point crossing, and that of $\overline{C_{HHH}^2}$ coming from modular invariance at genus-two. These formulas can be found in \cite{Collier:2019weq}, and together with \rref{result1sec4}, we have

\bea
\sqrt{\overline{C_{LLH}^2}} &\sim& 16^{-\frac{\Delta_H}{2}} e^{-\frac{S}{4}} \Delta_H^{2\Delta_L-\frac{c+1}{8}}  \\
\sqrt{\overline{C_{HHH}^2}} &\sim& \left(\frac{27}{16}\right)^{\frac{3\Delta_H}{2}}e^{-\frac{3S}{4}}\Delta_H^{\frac{5c-11}{72}} \\
\overline{C_{LLH}^3C_{HHH}}&\sim& \left(\frac{3\sqrt{3}}{16}\right)^{3\Delta_H}e^{-\frac{9 S}{4} }\Delta_H^{6\Delta_L-\frac{19+11c}{36}}  
\eea

Putting everything together we find
\begin{equation}
\frac{\overline{C_{LLH}^3C_{HHH}}}{\sqrt{\left(\overline{C_{LLH}^2}\right)^3\overline{C_{HHH}^2}}} \sim e^{-3S/4} \,,
\end{equation}
which is indeed exponentially suppressed confirming that non-Gaussianities are small. Note that the exponential terms in $\Delta_H$ as well as the power-law contributions that depend on $\Delta_L$ and $c$ exactly cancel out. In the end, this must come from the Virasoro crossing kernel, but we find it remarkable.

\subsection{Lorentzian limit}

Another fruitful avenue for future work would be to study the OPE in the Lorentzian limit. This will contain additional information about the OPE density. For example, by studying the Regge limit of the four-point function, one can obtain an OPE density weighted by a sine-squared function which suppresses the contributions near double-trace operators. The behavior of this density is then strongly affected by the Regge intercept of the CFT \cite{Caron-Huot:2020ouj}. In particular, there appears to be an important qualitative difference between theories that saturate the Regge-intercept bound and have $J_0=1$ where  the OPE coefficients can consistently be taken to be random, and theories with $J_0<1$ where the OPE coefficients must be peaked near double-trace operators and thus can not consistently be approximated as random variables. This suggests that not all chaotic CFTs are equal at the level of $C_{LLH}$ OPE coefficients, and in particular the statistical distribution of these variables depends on how chaotic a CFT really is.\footnote{It is worth emphasizing here that we are \textit{not} talking about the Regge limit relevant for holography and the bound on chaos \cite{Maldacena:2015waa}. Here, we are talking about the very deep Regge limit where Lorentzian times are parametrically bigger than any other timescale in the problem, in particular much greater than the scrambling time.}

We would like to understand how these statements generalize to higher moments of the OPE coefficients as we have explored them in this paper. For a six-point function, the additional cross ratios give a richer structure of Lorentzian time orderings (see for example \cite{Haehl:2017eob,Haehl:2017qfl,Haehl:2017pak,Haehl:2018izb,Anous:2019yku,Anous:2020vtw,Basu:2018akv,Chaudhuri:2018ymp,Roberts:2016hpo}) and it would be useful to uncover how the relevant Regge limits give additional insights into the OPE densities. We leave this for future work.

\subsection{Gravity interpretation}

Finally, it is worth speculating about the gravitational interpretation of our results. In the holographic dual, a $C_{LLH}$ OPE coefficient roughly estimates the probability of forming a black hole by colliding two light particles with high energy kinematics (see \cite{Das:2017cnv} for a related discussion). There are various interesting open questions remaining. In $d=2$ for example, why are the asymptotic formulas for quasi-primaries further exponentially suppressed  compared with those of Virasoro primaries? Moreover, we still lack a firm bulk understanding of the deep Lorentzian Regge limit, which probes physics beyond the OTOC saturation time. Is semi-classical gravity  capable of capturing some aspect of this process? 

Another avenue worthy of exploration is the connection between light-light-heavy OPE coefficients and wormholes. Wormholes cannot, under typical circumstances, be supported on positively curved surfaces, and here we are considering correlation functions on the plane. Could the addition of matter make it possible to support a wormhole \cite{Stanford:2020wkf,Garcia-Garcia:2020ttf}? And do wormholes give us new statistical information about $C_{LLH}$ OPE coefficients?

\section*{Acknowledgements}

We are happy to thank Shouvik Datta, Felix Haehl, Alex Maloney, Andrei Parnachev, Edgar Shaghoulian and Sasha Zhiboedov for fruitful discussions. AB is especially grateful to Matt Walters for stimulating discussions that inspired a part of this project. TA is supported by the Delta ITP consortium, a program of the Netherlands Organisation for Scientific Research (NWO) that is funded by the Dutch Ministry of Education, Culture and Science (OCW). JdB and DL are supported by the European Research Council under the European Unions Seventh Framework Programme (FP7/2007-2013), ERC Grant agreement ADG 834878.

\appendix

\section{Asymtotics of the Virasoro fusion kernel}\label{appendixfusion}

In this appendix, we collect some results regarding the asymptotic behaviour of the Virasoro fusion kernel. This kernel is built from the Barnes double gamma function $\gb(x)$; a meromorphic function with no zeros and poles at\footnote{throughout the appendix, $n$ and $m$ will be non-negative integers.} $x = -n b - m b^{-1}$. The function is defined via the recursion relation 
\begin{equation}
    \gb(x+b) := \frac{\sqrt{2\pi}b^{bx-\frac{1}{2}}}{\Gamma(bx)}\gb(x),
\end{equation}
the property that $\gb = \Gamma_{b^{-1}}$, and the normalization $\gb(Q/2) = 1$. 

The double sine function
\begin{equation}
    S_b(x) := \frac{\gb(x)}{\gb(Q-x)},
\end{equation}
is a meromorphic function with poles at $x= -m b -n b^{-1}$ and zeros at $x= Q + mb+nb^{-1}$. Many important properties of the Barnes gamma function and the double sine function, including several perturbative expansions, were explored and summarized in \cite{QuantumRegge}. 

The explicit expression for the fusion kernel involves a prefactor and a contour integral:

\begin{equation}
    \fker{s}{t}{1}{2}{3}{4} := P_b(P_i;P_s,P_t)P_b(P_i;-P_s,-P_t) \int_{C'}\frac{ds}{i}\prod_{k=1}^4 \frac{S_b(s+U_k)}{S_b(s+V_k)}.
\end{equation}
The prefactor $P_b$ is given in terms of $\Gamma_b$ functions
\begin{multline}
	P_b(P_i;P_s,P_t) = 
	\frac{\Gamma_b({\frac{Q}{2}}+i(P_s+P_3-P_4))\Gamma_b({\frac{Q}{2}}+i(P_s-P_3-P_4))}
	{\Gamma_b({\frac{Q}{2}}+i(P_t+P_1-P_4))\Gamma_b({\frac{Q}{2}}+i(P_t-P_1-P_4))}
	\\
	\times \frac{\Gamma_b({\frac{Q}{2}}+i(P_s+P_2-P_1))\Gamma_b({\frac{Q}{2}}+i(P_s+P_1+P_2))}
	{\Gamma_b({\frac{Q}{2}}+i(P_t+P_2-P_3))\Gamma_b({\frac{Q}{2}}+i(P_t+P_2+P_3))}
	\frac{\Gamma_b(Q+2iP_t)}{\Gamma_b(2iP_s)}.
\end{multline}
The arguments of the special function in the integrand are 
\begin{equation}
	\begin{split}
		U_1&=i(P_1-P_4)\\
		U_2&=-i(P_1+P_4) \\
		U_3&= i(P_2+P_3)\\
		U_4&=i(P_2-P_3)
	\end{split}
	\qquad
	\begin{split}
		V_1 &= Q/2+i(-P_s+P_2-P_4)\\
		V_2 &= Q/2+i(P_s+P_2-P_4) \\
		V_3 &= Q/2+iP_t \\
		V_4 &= Q/2-iP_t.
	\end{split}
\end{equation}
The integrand has four semi-infinite lines of poles to the left coming from the numerator, and four semi-infinite lines of poles to the right coming from the denominator. The contour $C'$ goes from $-i\infty$ to $i\infty$ in such a way that it passes in between the two families of poles. If the families overlap, we must deform the contour to the left or right and add the corresponding pole contributions. 

There are several known results regarding the asymptotic behaviour of the fusion kernel. We summarize some of these asymptotic formulas in Table \ref{asymptoticformulas}. To the best of our knowledge, all but the second formula were derived in the appendices of these papers \cite{Collier:2019weq,QuantumRegge,Belin:2021ryy,Esterlis:2016psv}. We will derive the second formula in the next section.

\begin{table}[t]
\centering
    \begin{tabular}{ll}\hline
        $\{P_sP_t;P_1P_2P_3P_4\}$ & Asymptotic expansion of $\log \bbf$ up to $\clo(1)$ terms   \\\hline 
        $\{HL;LLLL\}$ & $-2\log(4)(P+\delta_s)^2+ \pi(Q-2\alpha_t)P +2\left[-\frac{(3Q^2+1)}{4}+\sum_{i=1}^4 h_i\right] \log P\phantom{\bigg)}$\\[10pt]
        $\{HL;HLLL\}$ & $4(\delta_1-\delta_s)P \log 2 + \left(h_2+2h_3-2h_t-[\delta_1-\delta_s]^2\right)\log P$\\[10pt]
        $\{HL;HHLL\}$ & $\log\!\left(\frac{81\sqrt{3}}{16}\right)P^2\!+\!\left(\!-\frac{\pi Q}{2}\!+\![\delta_1\!+\!\delta_2]\log 27\! - \!\delta_s\!\log\!\frac{256}{27}\right)\!P\!+\!\left(\frac{1+7 Q^2}{6}-4h_t\right)\!\log\! P$\\[10pt]
        $\{HL;HLLH\}$ & $\big[h_2+h_3-h_t  -2(\delta_1-\delta_s)(\delta_4-\delta_s)\big]\log P$ \\[10pt]
        $\{HL;HHHH\}$ & \scalebox{0.9}{$3\!\log\!\left(\frac{27}{16}\right)\!P^2\!+\!\left[\!-\pi Q\! +\!(2\delta_s\!+\!\!\sum_{i=1}^4\! \delta_i)\!\log\!\frac{27}{16}\right]\!P\!+\!\!\left[\frac{5Q^2-1}{6}\!-\!2h_t\!+\!(\delta_2\!-\!\delta_3)^2\!+\!(\delta_1\!-\!\delta_4)^2\!\right]\!\log\! P$}\\[10pt]
        $\{HH;HHHH\}$ & $2(\delta_s-\delta_t)P \log\frac{27}{16}$  \\[5pt]\hline
    \end{tabular}
\caption{Asymptotic formulas for the logarithm of the Virasoro fusion kernel $\protect\fker{P_s}{P_t}{P_1}{P_2}{P_3}{P_4}$ up to $\clo(1)$ terms in the limit of large momenta with fixed differences. In the table, the label $H$ means that the corresponding momenta $P_i$ is heavy,  $P_i-\delta_i = P\rightarrow \infty$, while the $L$ variables and $\delta_i$ remain fixed.}
\label{asymptoticformulas}
\end{table}

\subsection{Asymptotics: three light external operators}
In this section we derive an asymptotic expression for the fusion kernel with $P_s - \delta_s = P_1-\delta_1 = P\rightarrow \infty$ and $P_t,P_2,P_3,P_4 \neq \bbi$. This is the result we need to find the asymptotics of the five-point function OPE density. Our strategy will be to first analyze the prefactor; we will look for zeros and poles. If the prefactor is finite and nonzero, we proceed to do an asymptotic expansion in the heavy momenta up to terms of order one. The last step is to study the integral.

In this limit, there are no poles in the prefactor and no zeros unless there is a fine-tuning of the type $i(P_2-P_3)=-\frac{Q}{2}- iP_t-nb-mb^{-1}$. These values of the momenta correspond to the \emph{Virasoro double-twist exchanges}. For these values, the integral contributes with a singularity, with a finite kernel in the limit. For now, it is safe to do an asymptotic expansion of the prefactor without paying too much attention to these contributions; we will comment on them at the end of the computation. For generic momenta, the asymptotic expansion of the prefactor reads:
\begin{multline}
        \log \left[P_b(P_i;P_s,P_t)P_b(P_i;-P_s,-P_t)\right]\\ =  \big(\pi Q + 2\pi i P_2 + 4[\delta_1-\delta_s]\log 2\big)P + \big(h_2 + 2 h_3-2h_t-[\delta_1-\delta_s]^2\big)\log P.
\end{multline}
Before picking a particular contour of integration, we will do an asymptotic expansion of the integrand. There are different regions because the asymptotic expansion of $S_b(x)$ is different in the lower and upper half-planes. The relevant terms in each region are
\begin{equation}
    \log \frac{S_b(s+U_k)}{S_b(s+V_k)} \sim
    \begin{cases}
    +2\pi i Q \sigma P & \Im\sigma>1\\
   +\pi Q P(i\sigma-1) & 0<\Im \sigma < 1\\
    -\pi Q P(i\sigma+1) & -1<\Im \sigma <0\\
    -2\pi i Q\sigma P & \Im \sigma <-1
    \end{cases}~. 
\end{equation}
This expression is parametrized in terms of $s \equiv \sigma P$. With a fixed value of $P$, we take the contour to be the imaginary axis and we integrate $\sigma$ from $-i\infty$ to $i\infty$. 
This integral is of the order of $\exp{-\pi Q P}$. The lines of poles to the left of the contour corresponding to $s = -i(P_2+P_3)-nb-mb^{-1}$ and $s= -i(P_2 - P_3)-nb-mb^{-1}$ cross the contour of integration. These contributions are dominated by the rightmost pole at $s =-U_3= -i(P_2+P_3)$ and have the following asymptotic behaviour
\begin{equation}
    \left(\log \frac{S_b(s+U_1)S_b(s+U_2)S_b(s+U_4)}{S_b(s+V_1)S_b(s+V_2)S_b(s+V_3)S_b(s+V_4)}\right)_{s=-U_3} =  - (\pi Q+2\pi i P_2) P +\clo(1).
\end{equation}
Thus, we find that
\begin{equation}
\label{fivepointkernel}
    \log \fker{s}{t}{1}{2}{3}{4} = 4(\delta_1-\delta_s)P\log 2+\big(h_2+2h_3-2h_t-[\delta_1-\delta_s]^2\big)\log P  + \clo(1).
\end{equation}
This result is also valid for light external operators even if they are fine-tuned to a Virasoro double-twist exchange. The reason is that the integral is dominated by the same pole that crosses the contour of integration. The zero in the prefactor and the singularity in the integral appear inside the $\clo(1)$ terms and do not couple to the heavy momenta. 

\subsection{Six-point asymptotics}\label{appendixSixpoint}
 Now we have all the technical tools to study the derivation of formula \rref{technicalsixpoint} and the result for the OPE statistics of the six-point function. From Table \ref{asymptoticformulas}, we have the following asymptotics for the crossing kernels:
 \begin{align}
\label{4pasym}
    \log \fker{P_2}{P_2'}{P_Y}{P_X}{P_X}{\tilde P_1}  
    &= -2 P_2^2 \log 4 + \pi(Q-2\alpha_{\clo_2'})P_2\\\nonumber & \eqspace{11em} + 2\left(-\frac{3Q^2+1}{4}+h_Y +2 h_X + \tilde h_1\right)\log P_2 + \ord(1),\\
    \log \fker{P_1}{\tilde P_1}{P_3}{P_2}{P_X}{P_W} &=  \left(-4 \log 2+\frac{9 \log 3}{2}\right)P^2+ \left(-\frac{\pi Q}{2} +[\delta_3+\delta_2]\log 27-\delta_1 \log \frac{256}{27}\right)P\\ \nonumber
    &\eqspace{16.5em} +\left(\frac{1+7Q^2}{6}-4\tilde h_1\right)\log P +\clo(1),
\end{align}
and a similar expression for the crossing kernel $\fker{P_3}{P_3'}{P_W}{P_Y}{\tilde P_1}{P_W}$.

From these expressions, we see that the crossing kernels favour the exchange of the identity operators $\clo_2',\clo_3'=\bbi$ with $\alpha_{\clo_2'},\alpha_{\clo_3'}=0$ in the sums over $\clo_2'$ and $\clo_3'$ in \rref{crossEq}. Subleading terms in the sum are suppressed by a factor of $\exp{-2\pi(\alpha_{\clo_2'}+\alpha_{\clo_3'})|P|}$. Hence, the leading contribution to the crossing equation has $\clo_2'=\clo_3'=\bbi$, this sets $\clo_1' = \bbi$ via the OPE coefficient $C_{\clo_1'\clo_2'\clo_3'}$. Having settled these variables, we find a delta function in the crossing kernel coming from the fusion\footnote{For a derivation of this formula, we recommend the appendices of \cite{Belin:2021ryy} and the footnotes of \cite{Collier:2019weq}.}
\begin{equation}
    \fker{\tilde P_1}{P'_1}{P'_3}{P_Y}{P_Y}{P_2'}_{P_1'=P_3'=P_2'=\bbi} = \delta(\tilde P_1 - P_Y)e^{\ord(1)}.
\end{equation}
This delta function allows us to evaluate the integral in the definition of the crossing kernel. After assembling the kernel, we find that 
\begin{equation}
\label{crossKer}
   \log \bbk_{\{P_1,P_2,P_3\},\{\bbi,\bbi,\bbi\}} \sim 3P^2 \log\left(\frac{3\sqrt{3}}{16}\right)+\frac{3}{2}\pi Q P+ \left[-\frac{5+11Q^2}{6}+4(h_W+h_X+h_Y)\right]\log P,
\end{equation}
where we have neglected the factor $-(\delta_1+\delta_2+\delta_3)P \log \frac{256}{27}$ because it is subleading.

\subsection{The five-point function}\label{appendixFivepoint}

\begin{figure}
    \centering
    \includegraphics{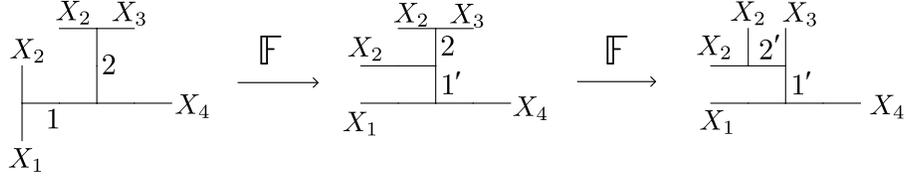}
    \caption{Sequence of crossing transformations that simplify the five-point function.}
    \label{fivepointrecursion}
\end{figure}

Figure \ref{fivepointrecursion} shows the sequence of moves we will be considering. In the heavy limit with fixed differences, the corresponding crossing equation reads
\begin{equation}
\label{fivepointOPE}
\begin{split}
 Q_5 &= \sum_{\clo_1'\clo_2'}\frac{C_{\clo_2'X_2X_2}C_{\clo_1'\clo_2'X_3}C_{\clo_1'X_1X_4}}{\den(P_1)\den(P_2)\den(\bar P_1)\den(\bar P_2)}
    \\ & \phantom{mmmm}\times \fker{P_1}{P_{\clo_1'}}{P_{X_4}}{P_2}{P_{X_2}}{P_{X_1}}
    \fker{P_2}{P_{\clo_2'}}{P_{X_3}}{P_{X_2}}{P_{X_2}}{P_{\clo_1'}}
    \fker{\bar P_1}{\bar P_{\clo_1'}}{\bar P_{X_4}}{\bar P_2}{\bar P_{X_2}}{\bar P_{X_1}}
    \fker{\bar P_2}{\bar P_{\clo_2'}}{\bar P_{X_3}}{\bar P_{X_2}}{\bar P_{X_2}}{\bar P_{\clo_1'}}
     \\
    &\sim \frac{C_{X_3X_1X_4}}{\den(P_1)\den(P_2)\den(\bar P_1)\den(\bar P_2)}\\ &\phantom{mmmm}\times\fker{P_1}{P_{X_3}}{P_{X_4}}{P_2}{P_{X_2}}{P_{X_1}}
    \fker{P_2}{\bbi}{P_{X_3}}{P_{X_2}}{P_{X_2}}{P_{X_3}}
    \fker{\bar P_1}{\bar P_{X_3}}{\bar P_{X_4}}{\bar P_2}{\bar P_{X_2}}{\bar P_{X_1}}
     \fker{\bar P_2}{\bbi}{\bar P_{X_3}}{\bar P_{X_2}}{\bar P_{X_2}}{\bar P_{X_3}}.
\end{split}
\end{equation}
In the last line, we used the exponential suppression of $P_2'$ to estimate the sum. The OPE coefficient $C_{\clo_1'\clo_2'X_3}$ sets $\clo_1'$ to $X_3$ whenever $\clo_2' = \bbi$. Corrections to this equation are exponentially suppressed in $P$ by the term $\exp{-2\pi \alpha_{\clo_2'}|P|}$.

The first fusion kernel has the following asymptotic behaviour (Eq. \ref{fivepointkernel})
\begin{equation}
    \log \fker{P_1}{P_{X_3}}{P_{X_4}}{P_2}{P_{X_2}}{P_{X_1}} = 4(\delta_2-\delta_1)P\log 2 + (h_{X_4}+2h_{X_2}-2h_{X_3}-[\delta_2-\delta_1]^2)\log P.
\end{equation}
Thus, we find the following the expression for the OPE coefficients
\begin{multline}
\label{G5asymptotics}
    \overline{C_{\clo_1X_1X_2}C_{\clo_2X_2X_3}C_{\clo_1\clo_2X_4}}\\ \sim C_{X_3X_1X_4}\;16^{-P^2} \exp{-3 \pi Q P +\left(-\frac{3Q^2+1}{2}+6h_{X_2}+2h_{X_3}+h_{X_4}-[\delta_2-\delta_1]^2\right)\log P}\times \textnormal{ (a.c.)}.
\end{multline}
Here,  we have dropped a factor $-4(\delta_1+\delta_2)P\log 2$, again because it is subleading. This time, the result is not universal; it explicitly depends on the OPE coefficient $C_{X_3X_1X_4}$.

\bibliographystyle{utphys}
\bibliography{asym}{}

\end{spacing}
\end{document}